\documentclass[fleqn,10pt]{wlscirep}
\usepackage[utf8]{inputenc}
\usepackage[T1]{fontenc}

\usepackage{graphicx}
\usepackage{dcolumn}
\usepackage{bm}
\usepackage{xcolor}
\usepackage{hyperref}
\usepackage{blkarray}
\usepackage{dsfont}

\title{Charge-parity switching effects and optimisation of transmon-qubit design parameters}

\author[1,2,*]{Miha Papi\v c}
\author[3]{Jani Tuorila}
\author[1]{Adrian Auer}
\author[1,2]{In\'es de Vega}
\author[1,+]{Amin Hosseinkhani}

\affil[1]{IQM, Georg-Brauchle-Ring 23-25,
80992 Munich, Germany}
\affil[2]{Department of Physics and Arnold Sommerfeld Center for Theoretical Physics, Ludwig-Maximilians-Universität München, Theresienstr. 37, 80333 Munich, Germany}
\affil[3]{IQM, Keilaranta 19, 02150 Espoo, Finland}

\affil[*]{miha.papic@meetiqm.com}
\affil[+]{amin.hosseinkhani@meetiqm.com}

\begin{abstract}
Enhancing the performance of noisy quantum processors requires improving our understanding of error mechanisms and the ways to overcome them. A judicious selection of qubit design parameters, guided by an accurate error model, plays a pivotal role in improving the performance of quantum processors. In this study, we identify optimal ranges for qubit design parameters, grounded in comprehensive noise modeling. To this end, we also analyze a previously unexplored error mechanism that can perturb two-qubit gates due to charge-parity switches caused by quasiparticles. Due to the utilization of the higher levels of a transmon, where the charge dispersion is significantly larger, a charge-parity switch will affect the conditional phase of the two-qubit gate. We derive an analytical expression for the infidelity of a diabatic controlled-Z gate and see effects of similar magnitude in adiabatic controlled-phase gates in the tunable coupler architecture. Moreover, we show that the effect of a charge-parity switch can be the dominant quasiparticle-related error source of a two-qubit gate. We also demonstrate that charge-parity switches induce a residual longitudinal interaction between qubits in a tunable-coupler circuit. We present a performance metric for quantum circuit execution, encompassing the fidelity and number of single and two-qubit gates in an algorithm, as well as the state preparation fidelity. This comprehensive metric, coupled with a detailed noise model, empowers us to determine an optimal range for the qubit design parameters. Substantiating our findings through exact numerical simulations, we establish that fabricating quantum chips within this optimal parameter range not only augments the performance metric but also ensures its continued improvement with the enhancement of individual qubit coherence properties. Our systematic analysis offers insights and serves as a guiding framework for the development of the next generation of transmon-based quantum processors.
\end{abstract}
\begin{document}

\flushbottom
\maketitle

\thispagestyle{empty}

\section*{Introduction}

While quantum processors continue to progress towards practical use, the errors present in current systems are still the most limiting factor. A dominant error in superconductor-based quantum computers is decoherence. 
There have been several proposals to mitigate it, either by designing new qubit types\cite{Gyenis_2021,Kjaergaard2020_review,Huang_2020} or by further optimizing the existing designs, typically focusing on increasing the coherence times of the circuit\cite{martinis2022,eun2022shape, Menke_2021}. However, in the latter case, one often encounters trade-offs between different circuit properties. For example in transmon qubits, which have emerged as the most widely-used qubit type in large-scale experiments\cite{arute_2019_supremacy,google_qec_2023,IBM_2023,cao_2023,wu_2021}, the suppression of charge noise comes at the cost of low anharmonicity which sets a lower bound on the duration of single-qubit operations\cite{Koch_2007}. 

This illustrates the importance of understanding the different errors affecting the quantum hardware, as well as the fact that an informed design of the circuit parameters must consider a plethora of possible error sources which are not necessarily limited only to the coherence properties of the circuit, but also include leakage errors in single qubit gates due to low anharmonicity, state preparation errors due to finite-temperature heating effects\cite{wenner_2013,jin_2015,heinsoo_2018}, as well as the parity-switching error presented in this manuscript. To elaborate further, only taking into account the coherence properties of the transmon and the low anharmonicity, one way to achieve better performance is to increase the transmon anharmonicity (in order to suppress potential leakage errors) while keeping the frequency of the transmon fixed. The latter condition, under the assumption of constant quality factors, ensures the coherence properties of the circuit remain unchanged. However this inadvertently leads into the regime where the transmon charge dispersion becomes more significant. One of the errors that is exponentially more pronounced in the low $E_J/E_C$ regime is related to the presence of electron-like excitations of the superconducting condensate, referred to as quasiparticles and the charge-parity of the transmon, thus prompting us to analyze these effects further so that a trade-off between the different error sources can be made.

Quasiparticles can be created through several mechanisms, and are known to cause different types of incoherent errors in superconducting qubit realizations \cite{Glazman_2021, Catelani_2022}. Particularly, quasiparticle tunnelling across the Josephson junction results in energy relaxation and dephasing in superconducting qubits\cite{Catelani_2011_PRL, Catelani_2011_prb, Corcoles_2011, Catelani_2012, Catelani_2014, Pop_2014}. 

\begin{figure}[t]
\centering
\includegraphics[width=.8\textwidth]{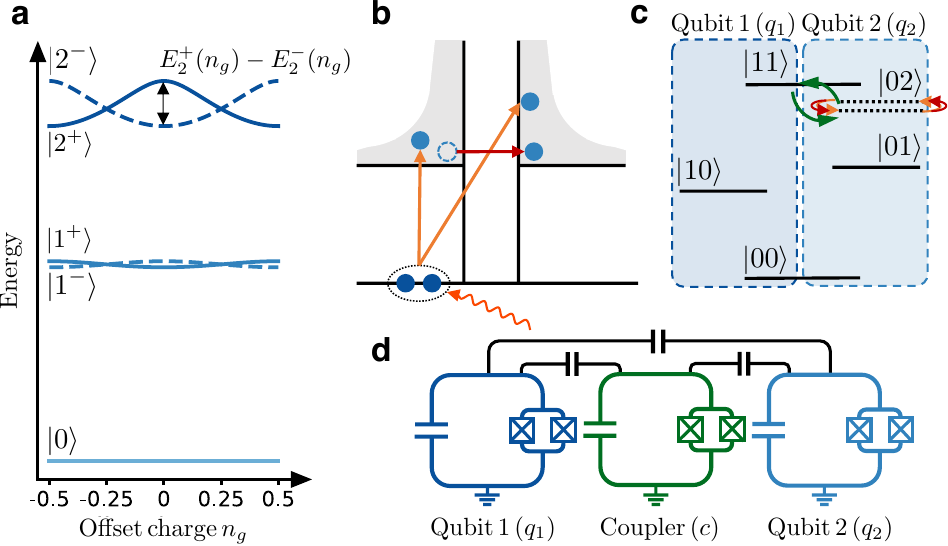}
\caption{\label{fig1:schematic} A schematic representation of the effect of a parity switch on a single transmon and on a two-qubit gate. \textbf{a} The energy diagram of a single offset-charge-sensitive transmon with $E_J/E_C = 10$ with two distinct parity manifolds, marked with + and -. While the ground state $|0\rangle$ also comprises of two distinct parity levels, the difference between them is not visible. \textbf{b} Illustration of two parity switching mechanisms. The vertical axis represents the energy, with the left and right regions corresponding to the two sides of the Josephson junction with the middle region corresponding to the insulator. The light grey area corresponds to the density of states of a BCS superconductor on both sides of the junction. \textit{Orange}: A high energy photon breaks a Cooper pair (dark blue) thus generating two quasiparticles (light blue), with one quasiparticle tunneling across the junction. \textit{Red}: A preexisting quasiparticle tunnels across the junction. \textbf{c} The energy level diagram of the states involved in the operation of a diabatic CPHASE gate. The Rabi oscillation between the levels $|11\rangle \leftrightarrow |02\rangle$ is marked with green arrows, however the larger charge dispersion of the second excited state means that both parity states of the $|02\rangle$ level cannot be considered degenerate anymore. \textbf{d} The lumped element model of the tunable-coupler circuit used in the implementation of diabatic CPHASE gates, consisting of two computational transmons (dark and light blue) referred to in the following as Qubits 1 and 2 (with indices $q_1$ and $q_2$), a flux-tunable coupler (green), denoted with the index $c$, and capacitive couplings between the transmons (black). The readout resonators and drive lines for the implementation of single-qubit gates are not included in the schematics.}
\end{figure}

Such detrimental quasiparticle-induced effects have, in turn, motivated research for finding mitigation strategies such as normal-metal traps \cite{Riwar_2016, Hosseinkhani_2017, Hosseinkhani_2018}, band-gap engineering \cite{Sun_2012, Riwar_2019, Marchegiani_2022} and improved qubit design \cite{Pan_2022}. There have also been efforts towards designing new types of superconducting qubits that are expected to be intrinsically robust against quasiparticle tunneling\cite{Smith_2020}. 

Suppression of charge-noise susceptibility of the transmon is achieved by adding a large shunt capacitor in parallel with a Josephson junction of a Cooper-pair box~\cite{Koch_2007}. However, the energy levels of the transmon exhibit a weak $2e$-periodic charge dispersion and are, thus, not completely independent of the offset charge on the transmon island. Since the presence of a quasiparticle shifts the island charge by $e$, the energy spectrum of the transmon can, thus, be divided into two distinct manifolds based on the parity of the number of quasiparticles on the island as shown in Fig.~\ref{fig1:schematic}a.

Switching of the charge-parity can occur either due to a preexisting quasiparticle tunneling across the Josephson junction on to the transmon island, or due to photon-assisted breaking of a Cooper pair, as depicted in Fig.~\ref{fig1:schematic}b\cite{Glazman_2021}. The timescale of these stochastic parity-switches is referred to as the parity-switching time and, as we argue in the Results section, it is typically much shorter compared to the time needed to obtain meaningful statistics from the quantum computer.

Since the difference between the two parity manifolds is strongly suppressed, quasiparticle effects in transmons have mainly been analyzed in the context of quasiparticle-induced decoherence. However, the parity-dependent energy splitting of the higher-excited states is much larger compared to that of the first-excited and ground states, as shown in Fig.~\ref{fig1:schematic}a. Therefore, parity-switching in the second excited state can potentially become a notable source of error for example in diabatic CZ (controlled-Z) or more generally CPHASE (controlled-phase) gates\cite{Strauch_2003}. There are different implementations of this gate scheme, most notably, statically coupled transmons\cite{Li_2019} and parametric drives\cite{Caldwell_2018}, tunable couplers and parametric drives\cite{Sete_2021} and the frequency-tunable coupler architecture\cite{Sung_2021,Xu_2021,Yan_2018,chu_2021,papic2023error,Marxer_2022,Sete_2021,Goto_2022,stehlik_2021_ibm_tqg,xu_2020}. Nonetheless, in all of these gate implementations, the second excited state of one transmon is populated during the gate operation, suggesting the gate could be susceptible to the charge-parity of one transmon in the system.

Besides the ability to perform fast and high-fidelity gate operations, one of the main reasons for the introduction of the tunable coupler is the fact that the static residual-ZZ interaction between the qubits can be completely suppressed by tuning the coupler transmon to a specific frequency\cite{Yan_2018,Sung_2021}, thus leading to reduced cross-talk and spectator qubit related decoherence\cite{Jurcevic_2022}. When determining this frequency, it has been shown that the level repulsion between the higher levels of the tunable coupler system are relevant\cite{chu_2021,Sung_2021}, thus implying that quasiparticle dynamics might affect our ability to effectively decouple the qubits. On the contrary, these level repulsions can serve as a means to enhance the ZZ interaction strength and facilitate the implementation of CPHASE gates entirely adiabatically, without populating the higher excited states \cite{collodo_2020}. Building upon this insight, we anticipate that parity switching may likewise exert a significant influence on the performance of adiabatic gate implementations.
    
In this paper, we develop an analytical theory of parity switches in a tunable coupler based architecture which is currently one of the most promising platforms for large-scale quantum computing\cite{arute_2019_supremacy,google_qec_2023,cao_2023,wu_2021}. We demonstrate that the effects of a parity switch in a two-qubit gate can be a relevant source of error, even in the transmon regime. Moreover, we show that this previously unidentified error can, in certain parameter regimes, be the dominant quasiparticle-induced error mechanism during a diabatic two-qubit gate, indicated by a comparison to currently achievable parity switching rates observed in superconducting circuits. Furthermore, we demonstrate that the inherent stochastic nature of parity-switching events limits the ability to suppress any unwanted longitudinal interactions between the qubits coupled through tunable couplers\cite{Mundada_2019}. We find that the magnitude of the unwanted interactions make this effect relevant as coherence times advance into the millisecond regime\cite{Wang_2022,place_2021}. 

Secondly, we introduce a systematic and versatile analysis aimed at identifying the optimal qubit design parameters. This comprehensive analysis involves accurately modeling dominant error sources and utilizing a performance metric tailored specifically for the quantum circuit under consideration. Importantly, we utilize analytical expressions for various infidelity terms within the performance metric, which significantly aids the scalability of our approach. By examining the performance metric's reliance on the transmon qubit design parameters, we delineate an optimal parameter range that maximizes overall performance. We demonstrate how our analysis and findings can offer invaluable guidance in the advancement of transmon-based quantum processors. Furthermore, since the established methodology in this paper is general-purpose, we anticipate its broader applicability across various hardware platforms.

\section*{Results}

\subsection*{Charge-Parity Modelling}
In this section, we first present how a parity switch affects a single transmon qubit. The Hamiltonian of an individual transmon, not taking into account potential higher order contributions to the Josephson energy \cite{willsch2023}, is given by \cite{Koch_2007,Serniak_2019}
\begin{equation}\label{eq:full_transmon_hamiltonian}
    \hat{H} = 4E_C\left(\hat{n} - n_g + \frac{P - 1}{4}\right)^2 - E_J \cos{\hat{\phi}},
\end{equation}
where the operator $\hat{n}$ represents the dimensionless charge and $\hat{\phi}$ is the superconducting phase operator across the Josephson junction. The variables $E_C$, $E_J$ and $n_g$ represent the charging energy of an electron (i.e. the energy required to add a single electron of the Cooper-pair to the transmon island), Josephson energy and dimensionless offset charge, respectively. The variables $\hat{n}$ and $\hat{\phi}$ are related via the canonical commutation relation $[\hat{\phi},\hat{n} ] = i$. Additionally we have included a discrete parity variable $P \in \{-1,+1\}$, corresponding to the parity of the number of \textit{electrons} that have tunneled across the junction. The parity term has the same effect as a shift of the offset charge by $\Delta n_g = 1/2$.

Denoting the eigenenergies of the original transmon Hamiltonian in Eq.~\ref{eq:full_transmon_hamiltonian} as $E_i$ with $i \in \{0,1,2,3,... \}$, the difference between the energy levels of the different parity states can be asymptotically approximated by \cite{Koch_2007}
\begin{equation}\label{eq:parity_difference}
    E_m^+(n_g) -  E_m^-(n_g) \simeq \epsilon_m \cos(2\pi n_g), 
\end{equation}
where the superscript refers to the parity and the charge dispersion $\epsilon_m$ is given by
\begin{equation}\label{eq:asymptotic_charge_dipersion}
    \epsilon_m \simeq (-1)^m E_C \frac{2^{4m + 5}}{m!}\sqrt{\frac{2}{\pi}}\left( \frac{E_J}{2 E_C}\right)^{\frac{m}{2} + \frac{3}{4}} e^{-\sqrt{8E_J /E_C}}.
\end{equation}
While the exponential suppression of the charge dispersion with the ratio $E_J/E_C$ is well-known and the main reason for the introduction of the transmon, the formula in Eq.~\ref{eq:asymptotic_charge_dipersion} also predicts a significant increase in the charge dispersion of higher excited states. This means that even though the effect of a parity switch may be small in the computational subspace, the effect can be significantly more pronounced if higher-excited states are involved in the operation of two-qubit gates. For example, we find $|\epsilon_2/\epsilon_1| \sim 40$ for $E_J/E_C \sim 50$. This difference is even more pronounced since certain effects scale with the square of the charge dispersion, as we show in the following. 

Consequently, the Hamiltonian of the single transmon in Eq.~\ref{eq:full_transmon_hamiltonian} can be approximated in the low-energy manifold and in the asymptotic limit of $E_J/E_C \gg 1$ as
\begin{equation}\label{eq:duffing_with_parity}
    \hat{H}/\hbar \simeq \left[
    \omega + \delta\omega(P,n_g) \right] 
    \hat{a}^\dagger \hat{a} + \frac{\alpha + \delta\alpha(P,n_g)}{2}\hat{a}^\dagger\hat{a}^\dagger \hat{a} \hat{a},
\end{equation}
where $\hat{a}$ are bosonic annihilation operators, and $\hbar\omega\equiv [E_1^+(n_g) + E_1^-(n_g)]/2 \simeq \sqrt{8 E_J E_C} - E_C$ and $\hbar\alpha \equiv [E_2^+(n_g) + E_2^-(n_g)]/2 \simeq -E_C$\cite{Catelani_2014} are the parity-averaged expressions for the transmon (angular) frequency and the anharmonicity in the asymptotic limit, respectively. Here, we have taken into account the fact that the different parities have almost identical parameters, and the small differences between them are taken into account with the two parameters depending on the parity, $ \delta\omega(P,n_g)$ and $\delta\alpha(P,n_g)$. 
The parity $P$ therefore divides the eigenstates of the Hamiltonian in Eq.~\ref{eq:full_transmon_hamiltonian} into two distinct manifolds, as illustrated in Fig.~\ref{fig1:schematic}a. Due to the rapid scaling of the charge dispersion shown in Eq.~\ref{eq:asymptotic_charge_dipersion}, we neglect in the following the effect of the parity switching on the first excited state, i.e. we set $\delta\omega = 0$, and only focus on the second excited state. Consequently, $\delta\alpha(P,n_g) = P\epsilon_2\cos(2\pi n_g)/(2\hbar)$.

\subsubsection*{The Diabatic $CPHASE$ Gate}\label{sec:diabatic_tqg}
We consider a non-adiabatic, i.e. diabatic, CPHASE gate based on the two-qubit gate scheme using tunable couplers that was analyzed in Refs. \cite{Sung_2021,Xu_2021,Yan_2018,chu_2021,papic2023error} with similar schemes proposed in Refs. \cite{Marxer_2022,Sete_2021,Goto_2022,stehlik_2021_ibm_tqg}. We show the circuit schematics of the tunable-coupler setup in Fig.~\ref{fig1:schematic}d. Here, the two computational transmons, which we refer to as Qubits 1 and 2 ($q_{1,2}$), are capacitively coupled with each other and to a third, frequency-tunable, transmon which is referred to as the tunable coupler or simply coupler ($c$). 

The main operation principle of the diabatic CPHASE gate is shown in Fig.~\ref{fig1:schematic}c. The CPHASE gate is implemented by tuning the frequency of the coupler closer to the frequency of the computational transmons by using a flux pulse. The conditional phase is collected during a Rabi oscillation between the $|11\rangle$ and $|02\rangle$ states of the computational qubits, as illustrated in Fig.~\ref{fig1:schematic}c. We model the circuit pictured in Fig.~\ref{fig1:schematic}d with the Hamiltonian \cite{Yan_2018,chu_2021}
\begin{align}\label{eq:tqg_ham}
    \hat{H}/\hbar &= \sum_{i \in \{q_1,c,q_2\}} \omega_i \hat{a}_i^\dagger \hat{a}_i + \frac{\alpha_i}{2} \hat{a}_i^\dagger\hat{a}_i^\dagger \hat{a}_i \hat{a}_i 
    - \sum_{\substack{i,j \in \{q_1,c,q_2\} \\ i\neq j}} g_{ij} (\hat{a}_i^\dagger - \hat{a}_i) (\hat{a}_j^\dagger - \hat{a}_j).
\end{align}

Since the second excited state of one of the computational transmons is significantly populated during the gate operation, the non-degeneracy of the two parity levels can have a direct effect, and therefore quasiparticle tunneling and photon-assisted pair breaking in the transmon can become a notable source of error in the gate operation. Further details about the numerical modeling and definition of the computational basis in relation to the states of the three constituent transmons are found in the Methods section.

\subsubsection*{Effective Model}
The tunable coupler circuit Hamiltonian in Eq.~\ref{eq:tqg_ham} is difficult to analyze and we must often rely on numerical studies\cite{chu_2021}. It is therefore beneficial to introduce an effective Hamiltonian that can approximate the physics of the system. 
Similar to Refs. \cite{heunisch2023tunable,Yan_2018}, we introduce the Schrieffer-Wolff transformation as a means to decouple the computational transmon states from the coupler states, and assume that the decoupled coupler remains in the ground state during the gate operations. Unlike the approach taken in Refs.~\cite{heunisch2023tunable,Yan_2018}, where the Hilbert space of the local transmons is truncated to the computational subspace, we also include the $|02\rangle$ state to account for the Rabi oscillation that is used to accumulate the conditional phase. More details of the Schrieffer-Wolff transformation can be found in the Methods section, where we show that the diabatic gate can be modelled with the following effective unitary

\begin{align}\label{eq:effective_unitary}
\hat{U}(t) \hat{=}
    \begin{blockarray}{ccccc}
    |00\rangle & |01\rangle & |10\rangle & |11\rangle & \\
    \begin{block}{(cccc)c}
      1 & 0 & 0 & 0  &\, |00\rangle \\
      0 & 1 &  0 & 0 &\, |01\rangle \\
      0 & 0 & 1 & 0 &\, |10\rangle \\
      0 & 0 & 0  & \sqrt{P_{11}} e^{i \phi(t)} &\, |11\rangle \\
    \end{block}
    \end{blockarray}.
\end{align}
Here, we have denoted the conditional phase with $\phi(t)$ and the population of the $|11\rangle$ state with $P_{11}$. Note that this operator is not necessarily trace-preserving, as part of the population of the $|11\rangle$ state might remain in the $|02\rangle$ state, due to potential calibration errors. More explicitly, $P_{11}$ is given by
\begin{equation}\label{eq:perturbative_P11}
    P_{11}(t) = 1 - \frac{2 \tilde{g}_{11,02}^2}{\Omega^2}\left[ 1 - \cos(\Omega t)\right],
\end{equation}
where we have defined the qubit-qubit detuning and the Rabi frequency of the $|11\rangle \leftrightarrow |02\rangle$ transition as $\tilde{\Delta} = \tilde{\omega}_{q_1} - \tilde{\omega}_{q_2} $ and $\Omega = \sqrt{(\tilde{\Delta} - \tilde{\alpha}_{q_2})^2 + 4\tilde{g}_{11,02}^2 }$, respectively. The parameters with the tilde denote the perturbed parameters of the original full Hamiltonian from Eq.~\ref{eq:tqg_ham}, which were derived by applying the Schrieffer-Wolff transformation (see Methods, more specifically Eqs.~\ref{eq:perturbative_hamiltonian_parameters_omega}-\ref{eq:perturbative_hamiltonian_parameters_g1102}). The conditional phase in Eq.~\ref{eq:effective_unitary} is given by
\begin{align}\label{eq:perturbative_cphase}
    \phi(t) &= \frac{1}{2}\left[ (\tilde{\alpha}_{q_2} - \tilde{\Delta})t + \pi \left(1 - \mathrm{sign}\{\cos(\Omega t/2) \} \right) \right] +  \mathrm{arctan}\left( \frac{\tilde{\Delta} - \tilde{\alpha}_{q_2}}{\Omega} \tan(\Omega t/2)\right).
\end{align}

Eqs.~\ref{eq:perturbative_P11} and \ref{eq:perturbative_cphase} can be used to assess the susceptibility of the gate parameters to a small perturbation, such as a parity switch. Due to the larger charge dispersion of the second excited state, and the fact that the second excited state of Qubit 2 ($q_2$) is populated during the gate operation, we can assume that the main contribution of the parity switch is the perturbation of the anharmonicity $\alpha_{q_2}$.

By treating the parity-dependent contribution to the anharmonicity $\delta\alpha$ from Eq. \ref{eq:duffing_with_parity} as a small perturbation, we can obtain the parity-dependent expressions for the conditional phase and $|11\rangle$ population
\begin{align}
    \phi(t_g,P_{q_2}) &\approx \phi_0 + \frac{\partial \phi}{\partial \alpha_{q_2}}\bigg|_{t = t_g} \delta\alpha_{q_2}(P_{q_2},n_g), \label{eq:cphase_taylor}\\
    P_{11}(t_g,P_{q_2}) &\approx 1 + \frac{1}{2}\frac{\partial^2 P_{11} }{\partial \alpha_{q_2}^2}\bigg|_{t = t_g} \left[ \delta\alpha_{q_2}(P_{q_2},n_g) \right]^2,\label{eq:Pee_taylor}
\end{align}
where the Taylor expansion of the optimal gate parameters for a small perturbation of $\alpha_{q_2}$ evaluated at the mean (parity averaged) anharmonicity $\alpha_{q_2}$ from Eq.~\ref{eq:duffing_with_parity} was employed. At this point, we stress again that $\delta\alpha(P,n_g) = P\epsilon_2\cos(2\pi n_g)/2$. While the above expressions are completely general also in the non-perturbative regime, the relations given in Eqs. \ref{eq:perturbative_P11} and \ref{eq:perturbative_cphase} can be used to obtain analytical expressions for ${\partial \phi}/{\partial \alpha_{q_2}}$ and ${\partial^2 P_{11} }/{\partial \alpha_{q_2}^2}$, which determine the susceptibility of the gate to charge-parity switches.

We realize from Eq.~\ref{eq:perturbative_P11} that the implementation of a high-fidelity gate with an arbitrary conditional phase $\phi_0$ requires $P_{11}(t_g)=1$, otherwise some population remains outside of the computational subspace. Therefore, the following condition for the gate time $t_g$ must hold: $\Omega t_g = n\cdot 2\pi$, in which $n \in \mathbb{N}$ is an integer number. This condition enables us to further simplify the relations for the susceptibility of the conditional phase to a parity switch, and we obtain up to the leading order,

\begin{equation}\label{eq:simplified_pert_cphase}
    \frac{\partial \phi}{\partial \alpha_{q_2}}\bigg|_{t = t_g} \approx  \frac{t_g}{2},
\end{equation}
and
\begin{equation}\label{eq:simplified_pert_P11}
     \frac{\partial^2 P_{11} }{\partial \alpha_{q_2}^2}\bigg|_{t = t_g} \sim
     \mathcal{O}\left(\frac{\tilde{g}_{11,02}^2}{(\omega_{q_2} + \alpha_{q_2} - \omega_{q_1})^4} \right).
\end{equation}
In the derivation of Eq.~\ref{eq:simplified_pert_cphase} we have neglected the terms proportional to $g_{q_1 c}g_{q_2 c}/(\omega_{q_2} - \omega_{c} + \alpha_{q_2})^3$ and $g^2_{q_2 c}/(\omega_{q_2} - \omega_{c} + \alpha_{q_2})^3$ and higher orders. Additionally Eq.~\ref{eq:simplified_pert_P11} only contains the lowest order scaling of the result. A list of all the assumptions used in the derivation of Eqs.~\ref{eq:simplified_pert_cphase} and \ref{eq:simplified_pert_P11} is included in the Methods section.

\subsection*{Gate Fidelity Limitations}
So far we have quantified how a parity switching event can affect the parameters of the gate unitary. In order to describe the gate performance in a quantum circuit, we also need to consider how frequently parity switching events occur.


Experiments of parity switching lifetimes typically show the parity switching time to lie in the broad range of $T_P \sim 100 \,\mu \mathrm{s}\,-\, 1 \,\mathrm{s}$ \cite{Diamond_2022,Riste_2013,Serniak_2018,Serniak_2019,kurter_2020,tennant_2022}. Even though the parity switching lifetime of transmons might increase in the future, e.g., due to better design and improved shielding, it appears that the parity lifetime may be fundamentally upper bounded by the high-energy quasiparticle burst events, which are observed to happen once every $10-50\,$s\cite{Thorbeck_2023,McEwen_2021,Cardani_2021,Gruenhaupt_2018}.

\subsubsection*{Kraus Operator Description}

Comparing the realistic range of parity lifetimes of superconducting qubits to the duration of a single two-qubit gate $t_g$, which is typically in the range of tens to hundreds of nanoseconds \cite{Marxer_2022,stehlik_2021_ibm_tqg,Sung_2021,Xu_2021,Rol_2019}, we observe that $t_g \ll T_P$. However, any meaningful application of a quantum computer will include re-running an algorithm, comprised of a large number of non-parallel gates $N_\mathrm{gates} $, in order to reduce the statistical uncertainty of the observable being evaluated. Denoting the number of runs by  $N_\mathrm{shots}$, the uncertainty of any observable generally scales as $1/\sqrt{N_\mathrm{shots}}$\cite{nielsen00} and therefore $N_\mathrm{shots}$ must be large. All together, the time needed to execute a full algorithm $T_\mathrm{alg}$ roughly scales as $T_\mathrm{alg} \propto N_\mathrm{shots} \cdot N_\mathrm{gates} \cdot t_g$, but it may be realistically even longer, due to the time needed to measure and reinitialize the quantum computer, and any possible pulse schedule compilation of the control electronics\cite{wack2021quality}.

As an example, in Refs.~\cite{arute_2019_supremacy,IBM_2023,vdBerg_IBM_PEC_2023} state of the art devices were run for a total of several minutes in order to obtain meaningful results, which implies the following clear separation of timescales
\begin{equation}\label{eq:separation_of_timescales}
    t_g \ll T_P \ll T_\mathrm{alg}.
\end{equation}
The left side of Eq.~\ref{eq:separation_of_timescales} indicates that the probability for a parity switch occurring during the operation of a single gate is very low, while the right-hand side suggests that a large number of parity switches can occur during an execution of an algorithm. This means that the effect of the charge-parity switch (CPS) on a diabatic CPHASE gate can be described by the following Kraus operators\cite{breuer_petruccione_oqs} acting on the two-qubit density matrix $\hat{\rho}$
\begin{equation}\label{eq:kraus_operators}
    \mathrm{CPHASE}_\mathrm{CPS}[\hat{\rho}] =\hat{U}_{-} \hat{\rho} \hat{U}^\dagger_{-} + \hat{U}_{ +} \hat{\rho} \hat{U}^\dagger_{ +},
\end{equation}
where $\hat{U}_{ \pm}$ are Kraus operators corresponding to the different parity implementations of the two-qubit gate. Eq.~\ref{eq:kraus_operators} can be interpreted as a stochastic application of two different gate operators; by assuming that the target conditional phase is $\phi_0$, they can be written as
\begin{equation}\label{eq:kraus_op_definition}
\hat{U}_\pm(t) = \frac{1}{\sqrt{2}}
\begin{pmatrix}
1 & 0 & 0 & 0 \\
0 & 1 & 0 & 0 \\
0 & 0 & 1 & 0 \\
0 & 0 & 0 & \sqrt{1 - \frac{\delta P_{11}}{4}} \,e^{ i\phi_0  \pm i \frac{\delta \phi}{2}}\\
\end{pmatrix},
\end{equation}
with $\delta \phi = \partial\phi/\partial \alpha_{q_2}\,\epsilon_2^{q_2}\cos(2\pi n_g^{q_2})/\hbar$ and $\delta P_{11} = \partial^2 P_{11}/\partial \alpha_{q_2}^2 \,\left[ \epsilon_2^{q_2} \cos(2\pi n_g^{q_2})/\hbar \right]^2/2$, which is a result of Eqs. \ref{eq:cphase_taylor} and \ref{eq:Pee_taylor}, together with $\delta\alpha(P,n_g) = P\epsilon_2\cos(2\pi n_g)/(2\hbar)$. We have additionally assumed that both parities are equally likely, however this assumption can also be easily relaxed. Note that the channel defined in Eq.~\ref{eq:kraus_operators} is not trace preserving when $\delta P_{11} > 0$. More specifically $\mathrm{tr}\{\mathrm{CPHASE}_\mathrm{CPS}[\hat{\rho}]\} \leq  \mathrm{tr}\{ \hat{\rho}\}$, where the equality holds when $\delta P_{11} = 0$. In this case the map is both completely positive and trace preserving and the completeness relation of the Kraus operators $\sum_{i=+,-} \hat{U}_i^\dagger \hat{U}_i = \mathbb{I}$ holds.

We have defined the Kraus operators in Eq.~\ref{eq:kraus_op_definition} so that there are small errors associated with each parity state. We show in the following that this corresponds to higher average gate fidelities, compared to having one parity state with a perfect fidelity and the second parity state with a larger error. This assumption is, therefore, equivalent to optimally calibrating the gate with respect to the parity switching error. 

\subsubsection*{Gate Fidelity}
Using the Kraus operator description in Eq.~\ref{eq:kraus_op_definition} allows us to make a formal analysis of performance metric of an arbitrary conditional phase gate in the presence of parity switches and resulting phase and leakage errors.

The average gate fidelity $\mathcal{F}$ in the presence of leakage can be computed by \cite{Emerson_2005,wood_2017}
\begin{equation}\label{eq:fidelity_definition}
    \mathcal{F}[\mathrm{CPHASE}_\mathrm{CPS}] = \frac{ \frac{1}{d}\sum_{i = +,-} \left|\mathrm{tr}\left\{\hat{U}_\mathrm{CPHASE}^\dagger\hat{U}_i\right\}\right|^2 + 1 - L}{d + 1},
\end{equation}
with the leakage parameter $L = 1 -  \mathrm{tr}\left\{\hat{U}_\mathrm{CPHASE}^\dagger \left[ \sum_{i = +,-} \hat{U}_i \hat{U}_i^\dagger\right] \hat{U}_\mathrm{CPHASE}\right\}/d$.
Here $\hat{U}_\mathrm{CPHASE}$ is the unitary operator of an ideal CPHASE gate and $d$ is the dimension of the computational Hilbert space, which in our case is $d = 4$.

The fidelity $\mathcal{F}$ of the operation given in Eq.~\ref{eq:kraus_operators} for small perturbations $\delta \phi$ and $\delta P_{11}$ is given by the following expression
\begin{align}
    \mathcal{F} \approx 1 - \frac{3}{80} \left[\frac{\partial \phi}{\partial \alpha_{q_2}}\bigg|_{t = t_g} \epsilon_2^{q_2} \cos(2\pi n_g^{q_2})/\hbar \right]^2 + \frac{1}{32} \frac{\partial^2 P_{11} }{\partial \alpha_{q_2}^2}\bigg|_{t = t_g} \left[ \epsilon_2^{q_2} \cos(2\pi n_g^{q_2})/\hbar \right]^2. \label{eq:cphase_Pee_infidelity}
\end{align}
Here, we observe that the effect on the fidelity is of second order in the charge dispersion, due to the coherent nature of a conditional phase error. However, we have shown in Eq.~\ref{eq:infidelity_scaling} of the Methods section that the infidelity of a series of $N$ gates is given by $1 -\mathcal{F}_N \approx  N(1 - \mathcal{F})$. This indicates that the error scales linearly with the number of gates, as is typical of incoherent errors, but quadratically in the error parameter $\delta\phi$, as is expected from a coherent error\cite{hashim_2021_rc,Wallman_2014}.

In deriving Eq.~\ref{eq:infidelity_scaling}, one can observe that the numerical prefactors in front of both terms (here $3/80$ and $1/32$) increase, and therefore the average gate fidelity decreases, if the gate is not calibrated in such a way that the error is equally distributed between both parity states. For this reason, we have chosen the Kraus operators as given in Eq.~\ref{eq:kraus_op_definition}.

We conclude from Eq.~\ref{eq:cphase_Pee_infidelity}, together with Eqs. \ref{eq:simplified_pert_cphase} and \ref{eq:simplified_pert_P11}, that the dominant effect of the parity switch event is the shift in the conditional phase, rather than leakage, since $({\partial \phi}/{\partial \alpha_{q_2}})^2 \gg {\partial^2 P_{11} }/{\partial \alpha_{q_2}^2}$ in the perturbative regime considered in this manuscript. Moreover the magnitude of the shift in conditional phase is given by Eq.~\ref{eq:cphase_taylor} and Eq.~\ref{eq:simplified_pert_cphase}. Further simplifying Eq. \ref{eq:infidelity_scaling} therefore results in
\begin{equation}\label{eq:simplified_infidelity}
    \mathcal{F} \approx 1 - \frac{3}{320} \left[ \epsilon_2^{q_2}t_g \cos(2\pi n_g^{q_2})/\hbar \right]^2.
\end{equation}

\subsection*{Numerical simulations}

Here, we compare the above results with the numerically exact treatment of full the Hamiltonian in Eq.~\ref{eq:tqg_ham}. The parity switch effect is taken into account in the numerical experiments by considering the transmon Hamiltonian from Eq.~\ref{eq:duffing_with_parity}. In other words, the system is simulated for all $2^3 = 8$ possible parity states and the results are averaged accordingly.

According to Eq.~\ref{eq:parity_difference}, the magnitude of the effect also depends on the offset voltage of Qubit 2, which is typically not known. Thus, we assume for simplicity that $n_g^{q_2} = 0$. Alternatively, as long as all the effects remain second order in the charge dispersion, one can also define an average charge dispersion as $\bar{\epsilon}_2^{q_2} = \epsilon_2^{q_2} \sqrt{ \int_0^1 \mathrm{d}n_g \cos^2(2\pi n_g) } = \epsilon_2^{q_2}/\sqrt{2} $.

\subsubsection*{Parity-Switch Induced Diabatic Gate Errors}\label{sec:gate_errors}

To achieve high-fidelity gate simulations, it is crucial to carefully select the Hamiltonian parameters in Eq.~\ref{eq:tqg_ham}. We have described how these parameters were chosen so that high-fidelity gates are possible with arbitrary ratios of $E_J/E_C$ of Qubit 2 in Table \ref{tab:tqg_pert_parameters} of the Methods section.
We also note that in general, we use parameters that closely resemble those in the implementation presented in Ref. \cite{Sung_2021}. We examine the effect of a parity-switch on the fidelity of a CZ gate and compare the perturbative analytical results to a full simulation of the Hamiltonian in Eq.~\ref{eq:tqg_ham}. The pulse shape used in the simulations (defined in Eq.~\ref{eq:flattop_gaussian}) is parameterized by its amplitude $A$, ramping up time $\tau_b$ and plateau duration $\tau_c$. The full duration of the flux pulse is therefore given by $T= \tau_c + 2\tau_b$.
\begin{figure}[t!]
\centering
\includegraphics[width=.9\textwidth]{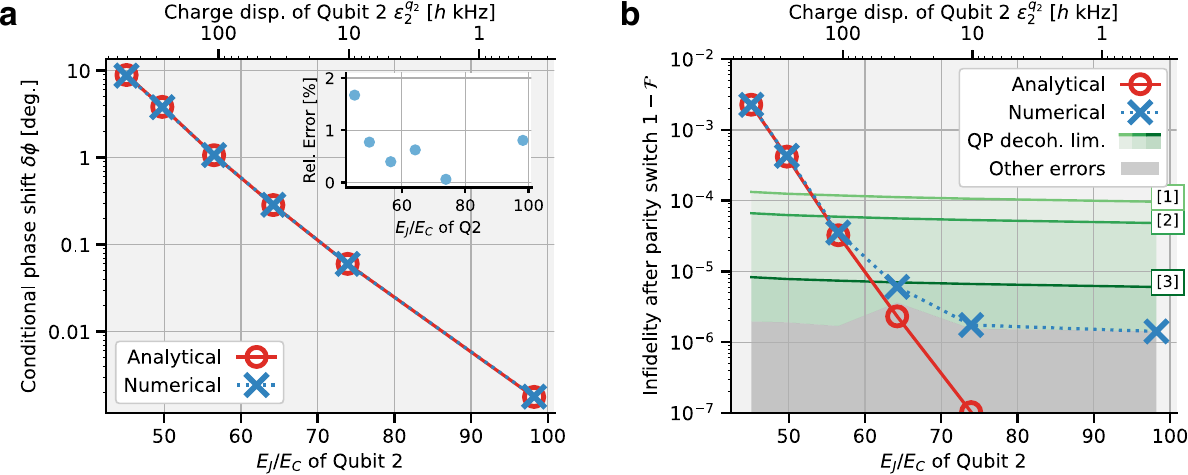}
\caption{\label{fig2:gate_results} Effects of a parity switch on the fidelity of a CZ gate. \textbf{a} Analytical predictions (red circles) of the conditional-phase difference of the CZ gate between the two parities. The theory predictions from Eqs. \ref{eq:simplified_pert_cphase} and \ref{eq:cphase_taylor} are compared to the phase extracted from a full numerical simulation of the gate (blue crosses), with the Hamiltonian parameters from Table \ref{tab:tqg_pert_parameters}. The inset displays the relative absolute error between the analytical predictions and the numerical results. The $x$-axis represents the $E_J/E_C$ ratio (bottom) and second excited state charge dispersion (top) of the qubit with the higher frequency (Qubit 2). The inset displays the relative error of the analytical prediction. \textbf{b} Gate infidelity after a parity switch on Qubit 2, where the theory predictions (red circles) only take into account the effect on the conditional phase, according to Eq. \ref{eq:simplified_infidelity}. The green lines are estimates of the upper bound for the infidelity contribution of quasiparticle induced decoherence on the two-qubit gate system, with the parity lifetimes $T_P$ measured by different references: [1] Risté, D., \textit{et al.} (2013)\cite{Riste_2013}, [2] Diamond, S., \textit{et al.} (2022)\cite{Diamond_2022} and [3] Kurter, C., \textit{et al.} (2022)\cite{kurter_2020}. More specifically, references [1] and [2] have reported the values $T_P = 1.25$ ms and $T_P = 2.5$ ms respectively. In reference [3], parity-switching times between 1 ms and 1.5 s were demonstrated. For this reference, we have used the value of  $T_P = 20$ ms, which approximately corresponds to the median parity-switching time of all the samples. The grey region is the region in which errors due to unwanted transitions during the gate operation are more prominent and the effect of a parity switch is negligible. It therefore represents the lower bound of the infidelity, which is largely independent of the ratio $E_J/E_C$.
}
\end{figure}

The conditional phase of the gate in the simulations is obtained from propagating the state $|\psi(t= 0)\rangle = \frac12 (1,1,1,1)^\mathrm{T}$ and extracting the conditional phase of the $|11\rangle$ state. Fig.~\ref{fig2:gate_results}a compares the analytical results in Eqs.~\ref{eq:cphase_taylor} and \ref{eq:simplified_pert_cphase} to the numerically obtained values and confirms that the two approaches agree up to a good accuracy. Since the magnitude of the charge dispersion in the numerical analysis is the same as in the analytical treatment, the error stems completely from the approximations made in evaluating $\partial \phi / \partial \alpha_{q_2}$. Furthermore, Fig.~\ref{fig2:gate_results}a also clearly shows that the parity-switching-induced shift in the conditional phase scales exponentially with the ratio of $E_J/E_C$. This is due to the scaling of the charge dispersion in Eq.~\ref{eq:asymptotic_charge_dipersion}. 

In Fig.~\ref{fig2:gate_results}b, we show numerical data of the full gate fidelity after the parity switch and, as a comparison, the corresponding analytical result calculated using Eq.~\ref{eq:simplified_infidelity}. The average gate fidelity in the numerical example is obtained from propagating a number of input states, reconstructing the effective superoperator of the gate from these simulations and subsequently using Eq.~\ref{eq:fidelity_definition} to obtain the average gate fidelity of the gate. We observe that the infidelity of the numerical simulation flattens for $E_J/E_C\gtrsim 75$, which is due to other errors in the gate implementation, such as leakage transitions during the pulse ramping up and down, which was confirmed by monitoring the population of the computational states during gate operation (see Supplementary Information for more details). In this region (in grey) the effect of a parity switch is not seen since it is too small compared to other errors. On the other hand at lower values of $E_J/E_C$, the numerical results overlap with the parity switching error predicted by the shift in the conditional phase. While clearly demonstrating the magnitude of the error, this result also shows that the leakage error contribution in Eq.~\ref{eq:cphase_Pee_infidelity} is negligible in the perturbative regime. The gate durations in Fig.~\ref{fig2:gate_results} are typically 45 ns $ \lesssim t_g  \lesssim $ 60 ns, with $\tau_c \sim t_g$.

Additionally, Fig.~\ref{fig2:gate_results}b compares the magnitude of the error due to quasiparticle related decoherence to the parity switch induced error described in this work. Since the quasiparticle induced characteristic decay times $T_1$ and $T_\phi$ depend on a large number of parameters and there are two possible quasiparticle generating mechanisms (see Fig.~\ref{fig1:schematic}b) \cite{Glazman_2021}, we only provide an upper bound based on the parity switching time observed in the references cited in the caption of Fig.~\ref{fig2:gate_results}. This upper bound is determined by noting that in the computational subspace\cite{Serniak_2018}
\begin{equation}
    \Gamma^{+-}_{00} + \Gamma^{+-}_{11} + \Gamma^{+-}_{01} + \Gamma^{+-}_{10} \approx 2/T_P,
\end{equation}
where the rates $\Gamma_{ij}^{+-}$ represent the transition rates between states $|i^\pm\rangle \rightarrow |j^\mp \rangle$ in different parity manifolds of a single transmon. Since amplitude damping noise has a larger effect on the fidelity compared to pure dephasing (see Table \ref{tab:noise_scaling}), we furthermore assume the worst-case scenario in which each quasiparticle-induced parity switching event results in a $T_1$ decay, so that $1/T_1^{qp} = \Gamma^{+-}_{01} + \Gamma^{+-}_{10} \approx 2/T_P$. This expression for $T_1^{qp}$, together with experimentally measured values of $T_P$, provides an approximate upper bound for the magnitude of the effect of the decoherence. Even though each green line in Fig.~\ref{fig2:gate_results}b uses a constant measured $T_P$, independent of $E_J/E_C$, the plotted infidelity contribution is not constant due to the varying gate duration $T = 2\tau_b + \tau_c$. 

We observe from Fig.~\ref{fig2:gate_results}b that depending on the ratio of $E_J/E_C$ and the parity lifetimes $T_P$, the contribution of parity switching to the infidelity of the two-qubit gate system can dominate the contribution from quasiparticle-induced decoherence. Note that the two-qubit gates are also typically the noisiest building blocks of a quantum algorithm \cite{hashim_2021_rc,arute_2019_supremacy,google_qec_2023,wu_2021}. We have also provided additional numerical results in the Supplementary Information, showing that the effect of a parity switch on the leakage becomes the main contribution to the infidelity at shorter gate times.

In order to compare the effect of the charge-parity switches to $1/f$-charge noise, which was previously thought to dominate the low $E_J/E_C$ regime, we included a comparison of the infidelity of a two-qubit gate due to $1/f$-type charge noise in the Supplementary Information, since the contribution of both of these error sources scales with the charge dispersion of the transmon. However, unlike the charge dispersion of the transmon which is maximized when $\cos(2\pi n_g) = 1$, the low-frequency charge noise decoherence rate is maximal at $\sin(2\pi n_g) = 1$ . This means that these two errors are mutually exclusive, i.e. if we were hypothetically able to tune $n_g$ to the value where the charge dispersion of the transmon is equal to zero, that point corresponds to the maximal decay rate due to low-frequency charge noise\cite{Koch_2007}. Nonetheless, in the comparison we have assumed in both the charge-parity switching analysis and the low-frequency charge noise analysis that we are at the noise hotspot, i.e. the value of $n_g$ where the effects are maximal, thus slightly overestimating both errors. We observe that at lower $E_J/E_C$ the charge-parity switching error is dominant and vice versa. The crossover between the infidelity due to the charge-noise-induced dephasing in the computational subspace and the parity switching exhibits a crossover at $E_J/E_C \sim 80$, which corresponds to a gate infidelity of $1-\mathcal{F} \sim 10^{-8}$, i.e. the charge-parity switching error is dominant for $E_J/E_C \lesssim 80$. However, due to the increased charge dispersion in the higher-excited levels of the transmon and the utilization of the second excited state, we have additionally analyzed the effect of the second excited state charge-noise-induced dephasing which was found to be much larger compared to the effects in the computational subspace. In this case, the crossover between the charge-parity error, which was again dominant at lower $E_J/E_C$, was found to occur at $E_J/E_C \sim 65$ and infidelities on the order of $1-\mathcal{F}\sim 10^{-6}$. We note here that our analysis overestimates the effect of the charge noise in a realistic scenario by neglecting the time correlations in the noise and should be treated as an upper bound of the magnitude of the effect. These results demonstrate that the parity-switching error is the dominant charge dispersion related error of the two-qubit gate in the regime of $E_J/E_C \lesssim 65$.

\subsection*{Parity-Switching Effects on the Adiabatic Interaction}\label{sec:idling_errors}
The tunable coupler architecture analyzed thus far can also be used to implement an \textit{adiabatic} CPHASE gate. In this case, the level repulsion between the states of the double-excitation manifold is used to obtain a phase shift of the $|11\rangle$ state, which accumulates the conditional phase with the rate
\begin{equation}\label{eq:zz_interaction}
    \zeta_\mathrm{ZZ} = \omega_{11} - \omega_{01} - \omega_{10} + \omega_{00},
\end{equation}
where $\hbar\omega_{ij}$ are the eigenenergies of the eigenstates which form the computational basis, as described in the Methods section. In this gate implementation the population ideally never leaves the computational subspace, however we will show that, also in the adiabatic implementation, the gate will suffer from charge-parity switching errors of a similar magnitude as in the diabatic case.

Additionally, one of the main benefits of the tunable coupler architecture is the ability to also suppress any residual interactions between the qubits during idling\cite{Mundada_2019,Yan_2018,chu_2021}, with unwanted ZZ coupling strengths demonstrated to be below 1 kHz \cite{Sung_2021}. In this section, we also show that uncontrolled parity-switching sets a lower bound on the minimum achievable unwanted ZZ interaction in such systems. 

In the presence of parity-switches in all three transmons of the tunable coupler system, the ZZ coupling rate of each of the distinct parity configurations will slightly differ. Since these differences are proportional to the charge dispersion, and therefore small, we can describe the parity-dependent ZZ coupling of this system with a first-order Taylor expansion, so that \begin{equation}\label{eq:parity_zz_coupling}
    \tilde{\zeta}_\mathrm{ZZ}(P_{q_1},P_c,P_{q_2} ) \simeq \zeta_\mathrm{ZZ}^0  + \sum_{i\in\{q_1 ,c, q_2 \}} \frac{P_i}{2} \left(\frac{\partial \zeta_\mathrm{ZZ}}{\partial \alpha_i} \epsilon_2^i +\frac{\partial \zeta_\mathrm{ZZ}}{\partial \omega_i} \epsilon_1^i \right),
\end{equation}
with $P_i \in \{-1,+1\} \, \forall i$. While we have still assumed $\epsilon_2 \gg \epsilon_1$, we have also acknowledged that the first-excited-state charge dispersion and the derivatives ${\partial \zeta_\mathrm{ZZ}}/{\partial \omega_i}$ can, in certain cases, be significantly larger compared to ${\partial \zeta_\mathrm{ZZ}}/{\partial \alpha_i}$. $\zeta_\mathrm{ZZ}$ therefore corresponds to the ZZ coupling strength of the system with parity-averaged parameters.

However, in order to evaluate the derivatives in Eq.~\ref{eq:parity_zz_coupling}, or the general value of the coupling rate $\zeta_\mathrm{ZZ}$ with fixed Hamiltonian parameters (without including parity effects), it is necessary to go to fourth-order perturbation theory in the coupling strengths, as was done in Refs. \cite{Sung_2021,chu_2021}. While the complete fourth-order expression is impractical for obtaining any analytical insight, by further assuming the hierarchy of the system parameters $\Sigma_{ic}\gg \Delta_{ic}\gg g_{ic} \gg g_{12}$, where $\Sigma_{ij} = \omega_i + \omega_j$ and $\Delta_{ij} = \omega_i - \omega_j$, for $i,j\in\{q_1,c,q_2 \}$, the cumbersome expressions can be significantly simplified into\cite{chu_2021} 
\begin{align}\label{eq:perturbative_zz_coupling}
    \zeta_\mathrm{ZZ} &\approx \frac{2\left[(\alpha_{q_1} + \alpha_{q_2})\tilde{g}_{01,10}^2 - 2\nu\tilde{g}_{01,10} (2\alpha_{q_1}\alpha_{q_2} + (\alpha_{q_1} - \alpha_{q_2})\Delta_{q_1 q_2}) \right]}{(\Delta_{q_1 q_2} + \alpha_{q_1})(\Delta_{q_1 q_2} - \alpha_{q_2})} + 2\nu^2\left[ 4\alpha_c + \frac{(\alpha_{q_1} + \alpha_{q_2})\Delta_{q_1 q_2}^2}{(\Delta_{q_1 q_2} + \alpha_{q_1}) (\Delta_{q_1 q_2} - \alpha_{q_2})}\right],
\end{align}
where $\nu = g_{q_1 c}g_{q_2 c}/(2\Delta_{q_1 c}\Delta_{q_2 c})\sim 10^{-3}$. 

\begin{figure}[t]
\centering
\includegraphics[width=.9\textwidth]{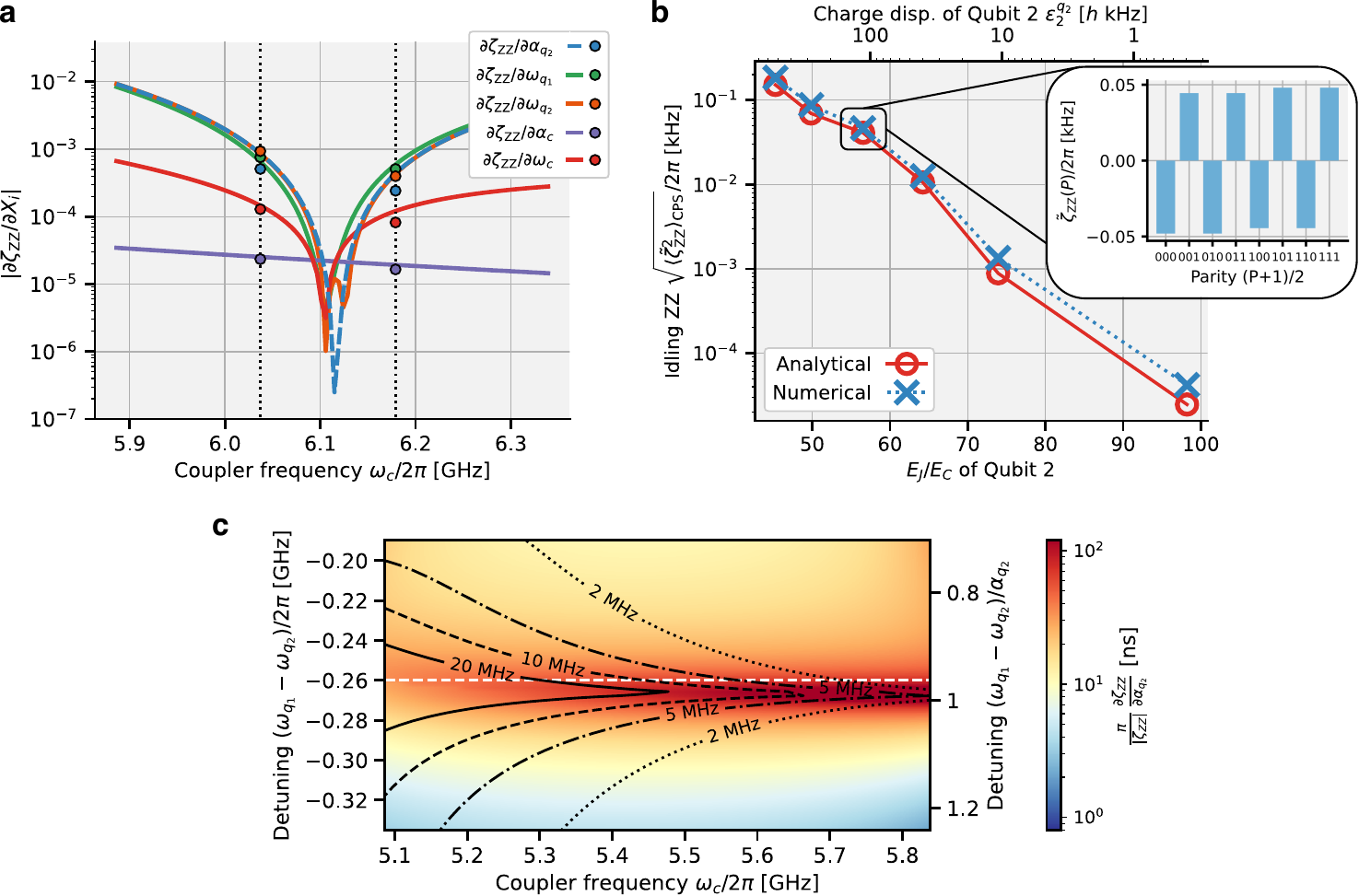}
\caption{\label{fig3:idling} Comparison of the analytical formulas with numerical results obtained via exact diagonalization in the idling regime. \textbf{a} Largest derivatives in Eq.~\ref{eq:parity_zz_coupling} obtained numerically (solid lines) and from Eq.~\ref{eq:perturbative_zz_coupling} (filled circles) using the parameters from Table \ref{tab:tqg_pert_parameters} and $\alpha_{q_2} = -270\cdot h$ MHz. The $x$-axis represents the coupler frequency, with the two idling frequencies denoted with black dashed vertical lines. Additional plots of the first-order derivatives at different qubit detunings are available in the Supplementary Information. \textbf{b} Comparison of numerical and analytical results for the root-mean-squared coupling strength at the idling point (defined in Eq.~\ref{eq:averaged_parity_zz}), with the bar graph displaying the numerical values of the coupling for the 8 different parity states for the highlighted data point. The parameters of the simulation are chosen identically as in Fig.~\ref{fig2:gate_results} and are listed in the Methods section, more specifically in Table \ref{tab:tqg_pert_parameters}. The parameters for the bar plot correspond to the parameters in panel a. \textbf{c} The susceptibility of an adiabatic CZ ($\phi_0 = \pi$) gate to a parity switch (defined in Eq.~\ref{eq:adiabatic_infidelity}) at different coupler frequencies and qubit detunings, measured in GHz (left axis) and in units of $\alpha_{q_2}$ (right axis), using the same parameters as in panel a. The black contours represent different values of $\zeta_\mathrm{ZZ}^0$, as indicated on the plot. The white dashed line indicates the detuning used in panel a.}
\end{figure}

Focusing on the idling regime, where $\zeta_\mathrm{ZZ}^0 = 0$, the difference between the derivatives obtained from Eq.~\ref{eq:perturbative_zz_coupling} and the numerically exact result is shown in Fig.~\ref{fig3:idling}a. We observe qualitatively good agreement in the vicinity of the two coupler idling frequencies of the system (vertical black dotted lines). 

Since the system has eight uncontrolled, rapidly (compared to experimental timescales, as in Eq.~\ref{eq:separation_of_timescales}) switching parity states, we define a parity-averaged idling interaction strength:
\begin{equation}\label{eq:averaged_parity_zz}
    \langle \tilde{\zeta}^2_\mathrm{ZZ} \rangle_\mathrm{CPS} = \frac{1}{2^3}\sum_{P_{q_1},P_c,P_{q_2} \in \{ -1, +1\}} \left[ \tilde{\zeta}_\mathrm{ZZ}(P_{q_1},P_c,P_{q_2}  ) \right]^2.
\end{equation}
This definition is the parity-averaged mean-square of the idling ZZ strength defined in Eq.~\ref{eq:parity_zz_coupling}. Parity-averaged coupling strengths are shown in Fig.~\ref{fig3:idling}b, where we observe a good agreement between the analytical result and the numerical data. The possible values of the quantity $\zeta_\mathrm{ZZ}(P_{q_1},P_c,P_{q_2}) $ for the third data point [also shown in panel a] are visualized in the bar graph. The bar graph demonstrates that Qubit 2 ($q_2$) is the dominant contributor to the perturbation, with small corrections due to the parity of Qubit 1 ($q_1$). This can be seen by noting that the residual coupling strength is dependent mostly on the parity of Qubit 2 ($q_2$). We have extended this analysis to larger values of $\Delta_{q_1q_2}$ in the Supplementary Information, to show that the parity-switching induced ZZ coupling strength is not always dependent on the parity of a single transmon in the system. The parity of the coupler in this set of parameters is largely irrelevant due to the high ratio of $E_{J_c}/E_{C_c} \approx 250$, but this is not always the case. For example, considering the gate implementation from Ref. \cite{Marxer_2022}, the ratio of $E_{J_c}/E_{C_c}$ for the coupler is much smaller and therefore, the coupler parity in this implementation has a much larger effect on the strength of the residual ZZ coupling.

The above results therefore present a fundamental limit on the magnitude of unwanted interactions that can be achieved in the tunable coupler architecture. However, the overall effect on an algorithm is more complex as it depends on the duration of the execution (since the interaction is always "on"), which in turn depends on the coherence times. We estimate that this error becomes relevant if the coherence times are of the order of $1/ \sqrt{\langle \tilde{\zeta}^2_\mathrm{ZZ} \rangle_\mathrm{CPS}}$, since a significant \textit{unwanted} conditional phase is accumulated if the algorithm duration $\Delta T$ (which is determined by the coherence times) is long enough such that $\Delta T \cdot\sqrt{\langle \tilde{\zeta}^2_\mathrm{ZZ} \rangle_\mathrm{CPS}} \sim 1$. For $E_J/E_C \sim 50$, the coherence time (or algorithm duration) needed to observe the parity-switching-induced residual coupling strength is on the order of 1 ms. Note that current fabrication processes are indeed already approaching this value\cite{Wang_2022,place_2021}.

Moving on the adiabatic CPHASE gate, we have observed in Fig.~\ref{fig3:idling}a,b that the state contributing most to the unwanted ZZ interaction in the idling regime is the second excited state of the high-frequency qubit, provided that $\omega_{q_1} - \omega_{q_2} \simeq \alpha_{q_2} $. This is due to the hybridization of the computational $|11\rangle$ state with the local transmon state $|0_{q_1}0_c 2_{q_2}\rangle$, which was also confirmed by the additional results included in the Supplementary Information. More importantly, this means that according to the same reasoning as in the diabatic gate case, the adiabatic gate can be described by the same pair of Kraus operators as defined in Eq.~\ref{eq:kraus_op_definition}, and since we are assuming adiabaticity, we can automatically set $\delta P_{11} = 0$. Again, applying the same reasoning as in the diabatic case, the fidelity of the adiabatic gate is given by 
\begin{equation}\label{eq:adiabatic_infidelity}
    \mathcal{F} \approx 1 - \frac{3}{80} \left[\frac{\partial \phi}{\partial \alpha_{q_2}}\bigg|_{t = t_g} \epsilon_2^{q_2} \cos(2\pi n_g^{q_2})/\hbar \right]^2 = 1 - \frac{3}{80} \left[\underbrace{\frac{\partial \zeta_\mathrm{ZZ}}{\partial \alpha_{q_2}} \frac{\phi_0}{|\zeta_\mathrm{ZZ}^0 |}}_\text{Fig.\,\ref{fig3:idling}c} \epsilon_2^{q_2} \cos(2\pi n_g^{q_2})/\hbar \right]^2.
\end{equation}
In the second step we have again assumed a flat pulse with duration $t_g$ and neglected the ramping up and down phases of the flux pulse. With this simplification, the conditional phase is given by $\phi_0 = \zeta_\mathrm{ZZ}^0 t_g$. Contrasting the expressions in Eq.~\ref{eq:adiabatic_infidelity} and Eq.~\ref{eq:simplified_infidelity}, we can see that if the quantity $\partial \zeta_\mathrm{ZZ}/\partial \alpha_{q_2}\cdot \phi_0 /|\zeta_\mathrm{ZZ}^0 |$ is on the order of $\sim 10\,$ns, the error of the adiabatic gate is similar to the diabatic case (displayed in Fig.~\ref{fig2:gate_results}). More specifically, the conditional phase shift due to a charge-parity switch is given by $\delta\phi_d \approx t_g/2 \cdot \epsilon_2^{q_2}/\hbar$ in the diabatic case (according to Eq.~\ref{eq:simplified_pert_cphase}) and by $\delta\phi_a \approx \partial \zeta_\mathrm{ZZ}/\partial \alpha_{q_2}\cdot \phi_0 /|\zeta_\mathrm{ZZ}^0 | \epsilon_2^{q_2}/\hbar$ in the adiabatic case (ignoring the factor $\cos(2\pi n_g^{q_2})$).

The quantity $\partial \zeta_\mathrm{ZZ}/\partial \alpha_{q_2}\cdot \phi_0 /|\zeta_\mathrm{ZZ}^0 |$ is  numerically analyzed in Fig.~\ref{fig3:idling}c, where we can see that in the vicinity of the horizontal line where $\omega_{q_1} - \omega_{q_2} = \alpha_{q_2}$, the hybridization with the second excited state results in an increased sensitivity to the charge-parity switches. However, the exact value of $\partial \zeta_\mathrm{ZZ}/\partial \alpha_{q_2}\cdot \phi_0 /|\zeta_\mathrm{ZZ}^0 |$ is very dependent on the value of the qubit-qubit detuning, but generally the regime where  $\omega_{q_1} - \omega_{q_2} < \alpha_{q_2}$ is more favorable. In the regime where $\omega_{q_1} - \omega_{q_2} \approx \alpha_{q_2}$, the observed effect of a charge-parity switch on the fidelity (which is proportional to the square of the plotted values according to Eq.~\ref{eq:adiabatic_infidelity}) can be an order of magnitude larger than in the diabatic case. Conversely, in the regime where $\omega_{q_1} - \omega_{q_2} < \alpha_{q_2}$, the infidelity is up to two orders of magnitude smaller. More specifically, at $E_J/E_C = 50$ (corresponding to $\epsilon_2^{q_2} \approx 250\,h\,$kHz), the maximal fidelity achievable in the range of the plot in Fig.~\ref{fig3:idling}c is $\mathcal{F} \sim 1 - 10^{-7}$, and the minimal fidelity is $\mathcal{F} \sim 1 - 10^{-2}$.

\subsection*{Optimal Qubit Parameters}

\begin{figure}[h!]
\centering
\includegraphics[width=.9\textwidth]{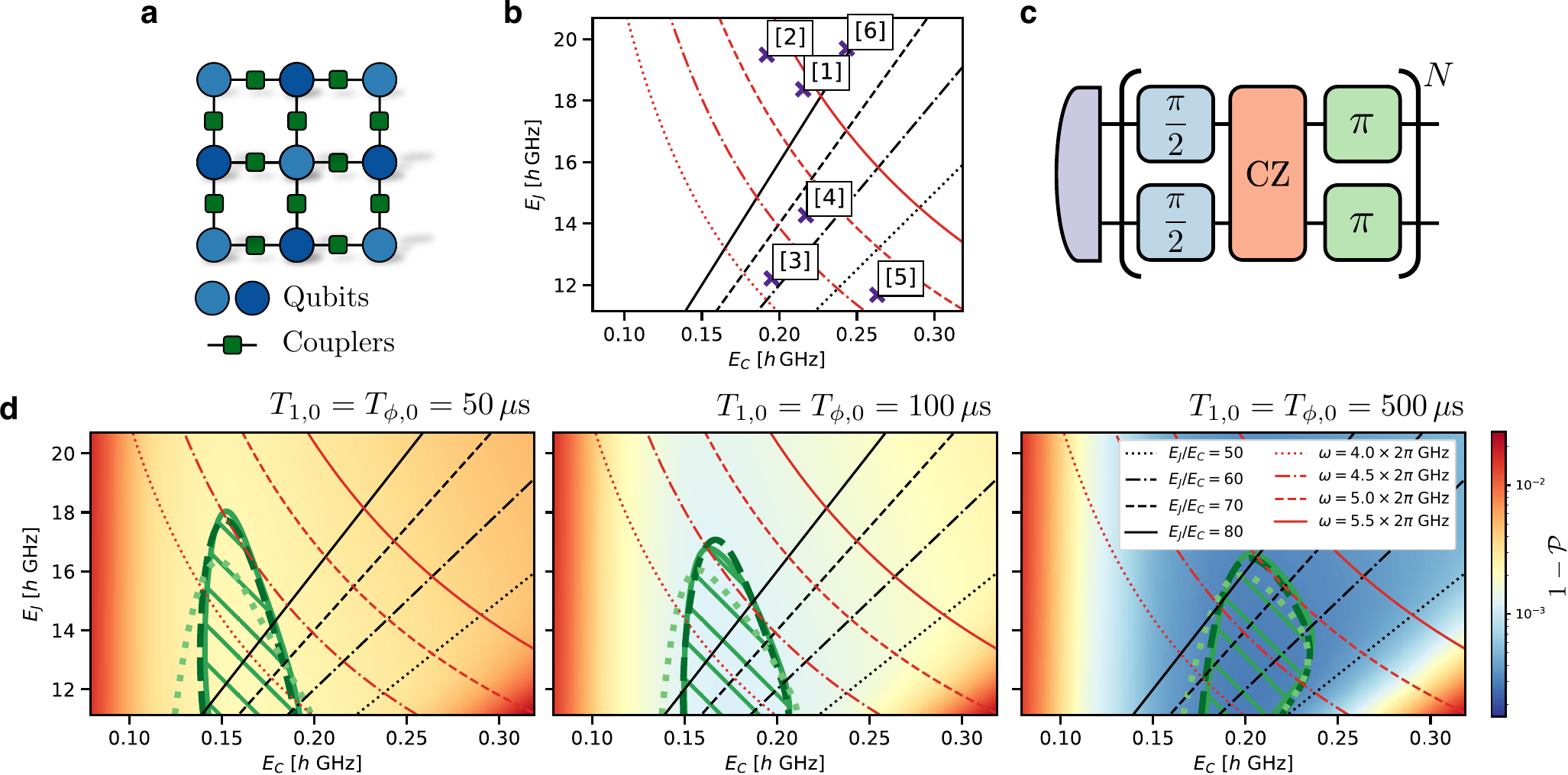}
\caption{\label{fig4:opt_EjEc} Determining optimal transmon parameters. \textbf{a} Schematic representation of the square grid architecture, with high- (dark blue) and low-frequency (light blue) transmons, connected via tunable couplers (green). \textbf{b} Values of $E_J$ and $E_C$ of Qubit 2 ($q_2$) from different experimental implementations of the tunable coupler transmon architecture. Both parameters are extracted from the reported $\omega_{q_2}$ and $\alpha_{q_2}$, and therefore the points are only an approximation\cite{willsch2023}. The annotations refer to the following references [1] Collodo, M. C., \textit{et al.} (2020)\cite{collodo_2020}, [2] Xu, Y., \textit{et al.} (2020)\cite{xu_2020} [3] Sung, Y., \textit{et al.} (2021)\cite{Sung_2021}, [4] Wu, Y., \textit{et al.} (2021)\cite{wu_2021}, [5] Xu, H., \textit{et al.} (2021)\cite{Xu_2021} and [6] Google Quantum AI (2022)\cite{google_qec_2023}. The black lines represent constant ratios of $E_J/E_C$ and the red lines correspond to contours of constant qubit frequency $\omega_{q_2}$. \textbf{c} Schematic representation of the circuit used to infer the weights in Eqs.~\ref{eq:total_infidelity} and \ref{eq:total_infidelity_full}, with the state preparation pictured on the left (in purple), single-qubit $\pi$ and $\pi/2$-rotations in green and blue respectively, and the two-qubit CZ gate in orange. $N$ is the number of times the pictured circuit (without the state preparation) is repeated before measurement and therefore an integer determined by the reference coherence time. In our case, we consider $N = \left\lfloor \frac{T_{1,0}}{10 ( t_\mathrm{TQG} + 2t_\mathrm{SQG}) } \right\rfloor$. \textbf{d} The function $1 - \mathcal{P}$ defined in Eq.~\ref{eq:total_infidelity_full} plotted for different values of the second (higher frequency) transmon $E_J$ and $E_C$. We consider a single-qubit gate implemented with a Gaussian DRAG pulse with a duration of $t_\mathrm{SQG} = 16\,$ns, a two-qubit gate duration of $t_\mathrm{TQG} = 50\,$ns and three different reference coherence times indicated on top of each panel. For all three cases, the reference $T_{\phi,0} = T_{1,0}$ at $E_J = 12\cdot h\,\mathrm{GHz}$ and $E_C = 0.2\cdot h\,\mathrm{GHz}$ for Qubit 2 ($q_2$) with the parameters of Qubit 1 ($q_1$) given in Table \ref{tab:tqg_pert_parameters}. All the parameters (transmon parameters and decay times) are scaled accordingly to different values of $E_J$ and $E_C$, and for each qubit individually, as described in Tables~\ref{tab:tqg_pert_parameters} and \ref{tab:noise_scaling}. The striped green area marks the region in the plot with the lowest values of $1 - \mathcal{P}$, defined by the 10th percentile of the plotted values. The darker green contour is obtained with a density matrix simulation of the circuit from panel c with the same errors, but instead of evaluating the function $1-\mathcal{P}$, it is obtained by minimizing the infidelity of the state before measurement. The lighter green dotted contour is the optimal region obtained using the advanced noise model from Table~\ref{tab:noise_scaling}.}
\end{figure} 

Having established the magnitude of the parity-induced error on a two-qubit gate, we have shown that this error can be mitigated by increasing the $E_J/E_C$ ratio of the transmon. However, there are other error sources present in such architectures\cite{papic2023error}, and while increasing the $E_J/E_C$ ratio will suppress the parity-switching errors, it may also increase the contribution of other possible error sources. Therefore, in order to find better parameters for future transmon-based quantum computers, we must evaluate the contributions of all errors affecting the system. In particular, we estimate optimal regions for the qubit parameters $E_J$ and $E_C$, where the errors contributing to the gate and state preparation infidelities are minimized.

We consider a number of different error sources relevant to superconducting qubits:
\begin{enumerate}
    \item $T_1$ decay due to the coupling to a bath of two-level systems \cite{siddiqi_2021_review,Muller_2019,carroll_2022,Premkumar2021,cho2022,Lisenfeld_2019,abdurakhimov_2022}.
    \item $T_\phi$ pure dephasing due to the coupling to magnetic flux noise \cite{rower_2023,Braumuller_2020,Yan_2016,siddiqi_2021_review}.
    \item Leakage affecting single-qubit gates due to low anharmonicity \cite{motzoi_2009}.
    \item State preparation errors due to finite-temperature heating effects \cite{wenner_2013,jin_2015,heinsoo_2018}, without the presence of active reset.
    \item Errors in the two-qubit gate operation due to parity switch effects that are analyzed in this manuscript.
\end{enumerate}
In Methods, we show how the above error sources scale with the transmon Hamiltonian parameters $E_J$ and $E_C$. We have not included any errors related to the control and calibration of the individual gates, as such errors do not explicitly depend on the qubit parameters and their inclusion, therefore, would not significantly alter the presented results. Similarly, measurement errors are present, but have no explicit dependence on $E_J$ and $E_C$.


For simplicity, we consider the same tunable coupler system as in the previous section, but arranged in a square grid, as pictured in Fig.~\ref{fig4:opt_EjEc}a. In order to avoid frequency crowding issues with such a connectivity\cite{berke_2022}, the qubits in the array are divided into low and high-frequency transmons. Furthermore, to prevent next-nearest-neighbor frequency collisions, even qubits assigned to the high- or low-frequency groups must possess non-identical frequencies, as demonstrated in Ref~\cite{hertzberg_2021}. Nonetheless, the frequency spread within each group is considerably smaller compared to the disparity between the means of these two groups. While our analysis focuses on a single qubit pair, extending the results to encompass a parameter distribution is straightforward. Additionally, we will define a broader region of optimal parameters spanning several hundred MHz in frequency, ensuring that all transmons within the square grid can fit within this designated range.

In the following, we consider two noise models. It has been demonstrated that the gate infidelity due to decoherence is, up to first order, independent of the unitary dynamics\cite{abad_2022}. Therefore, in the basic noise model, presented in Table~\ref{tab:noise_scaling}, both $T_1$ and $T_2$ decay are modeled by applying the corresponding Kraus operators \textit{after} the gate operation. We further assume that the detuning between the computational transmons is consistently $\omega_{q_1} - \omega_{q_2} \approx \alpha_{q_2}$, and the anharmonicity of the qubits is similar. Given these conditions, we parameterize the entire system in terms of $E_J$ and $E_C$ of the higher-frequency transmon. Examples of parameter values, obtained from experimental demonstrations of CZ gates, are plotted in Fig.~\ref{fig4:opt_EjEc}b.

In the advanced noise model (see Table~\ref{tab:noise_scaling}), we account for the population leaving the computational subspace during the operation of the diabatic CZ gate \cite{abad2024}. Additionally, in the advanced noise model we also consider the qubits to be detuned by $\alpha_{q_1}/2 \approx \alpha_{q_2}/2$ during the execution of single-qubit gates (in order to prevent driving unwanted leakage transitions)\cite{marxer_aps_2024}. The low-frequency qubit is then tuned so that $\omega_{q_1} - \omega_{q_2} = \alpha_{q_2}$ only during the two-qubit gate operation, thus increasing the flux-noise susceptibility during a CZ gate, according to Eq.~\ref{eq:flux_dipsersion}.

\begin{figure}[t]
\centering
\includegraphics[width=.8\textwidth]{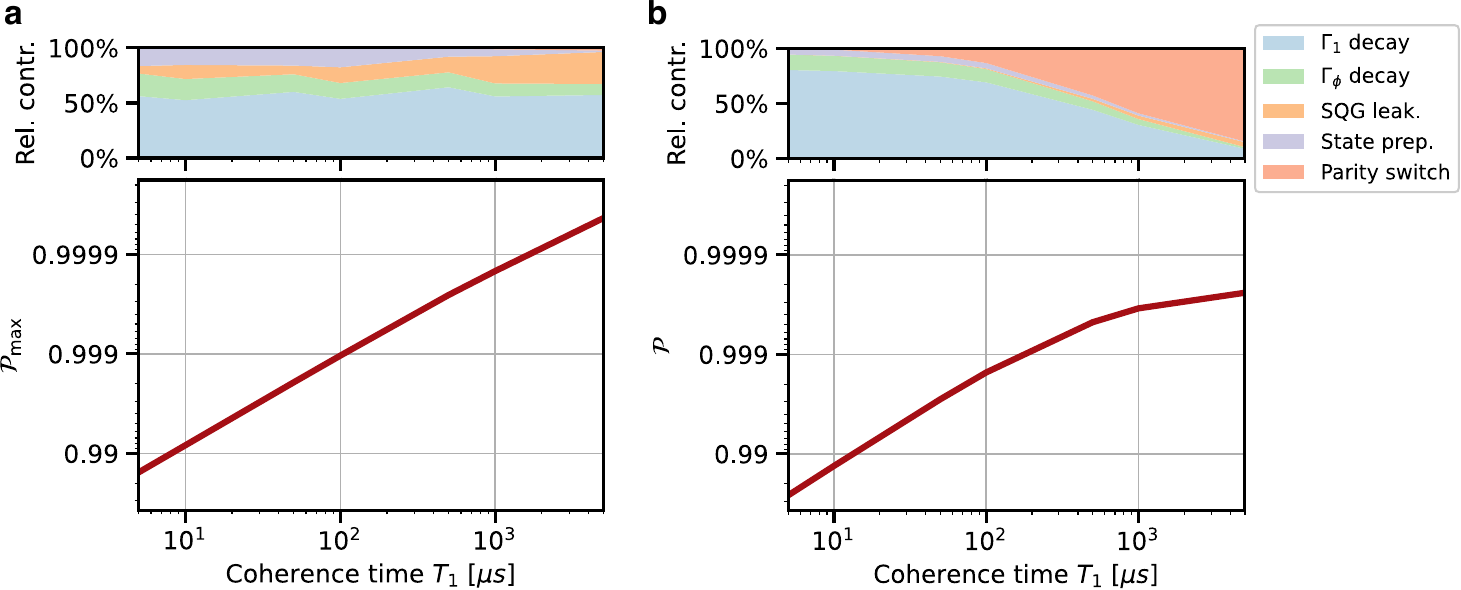}
\caption{\label{fig5:example_opt_EjEc} The relative contributions to the value of $1 - \mathcal{P}$ (in Eq.~\ref{eq:total_infidelity}) from the error sources listed in Table \ref{tab:noise_scaling} (top) and the value of the performance metric $\mathcal{P}$ as a function of coherence times with assuming $T_{\phi,0} = T_{1,0}$ at the reference point (bottom). Here we considered the same reference point as used in Fig.~\ref{fig4:opt_EjEc}d. \textbf{a} A well-designed system, where the parameters $E_J$ and $E_C$ are optimally adapted so that $\mathcal{P}$ is maximized for each value of the coherence time on the $x$-axis. \textbf{b} $\mathcal{P}$ at various coherence times while keeping $E_J$ and $E_C$ fixed, corresponding to the values marked as point [5] in Fig.~\ref{fig4:opt_EjEc}b. }
\end{figure} 

In order to quantify the performance of an algorithm execution with a specified pair of parameters $E_J$ and $E_C$ in mind, we define a performance metric $\mathcal{P}$ which we will then maximize. We further define  $\mathcal{P}$ as one minus a weighted sum of the infidelity contributions of all the relevant errors listed in Table \ref{tab:noise_scaling}. This sum can be written as
\begin{equation}\label{eq:total_infidelity}
    1 - \mathcal{P} = \sum_{i =T_1, T_\phi, \mathrm{parity} } w_{\mathrm{TQG},i} (1 - \mathcal{F}_{\mathrm{TQG},i}) + \sum_{i = q_1,q_2} \sum_{j = T_1, T_\phi, \mathrm{leak.}} w_{\mathrm{SQG},i,j} (1 - \mathcal{F}_{\mathrm{SQG},i,j}) + \sum_{i = q_1,q_2}  w_{\mathrm{SP},i} (1 - \mathcal{F}_{\mathrm{SP},i}),
\end{equation}
where the summation runs across all infidelity contributions, or more explicitly for the simple algorithm pictured in Fig.~\ref{fig4:opt_EjEc}c and the basic noise model from Table~\ref{tab:noise_scaling}
\begin{align}\label{eq:total_infidelity_full}
    1 - \mathcal{P} &=  \frac{2}{5}(\Gamma_1^{q_1} + \Gamma_1^{q_2} ) t_\mathrm{TQG} + \frac{1}{5}(\Gamma_\phi^{q_1} + \Gamma_\phi^{q_2} ) t_\mathrm{TQG} + \frac{3}{80}\left(\frac{t_\mathrm{TQG}}{2\hbar}\epsilon^{q_2}_2 \right)^2 \nonumber \\
    &+ 2\left[\frac{1}{3}\Gamma_1^{q_1} t_\mathrm{SQG} +  \frac{1}{6}\Gamma_\phi^{q_1} t_\mathrm{SQG} \right] +  \frac{1}{3}P_\mathrm{leak.}^{q_1}+ 2\left[\frac{1}{3}\Gamma_1^{q_2} t_\mathrm{SQG} +  \frac{1}{6}\Gamma_\phi^{q_2} t_\mathrm{SQG} \right] +  \frac{1}{3}P_\mathrm{leak.}^{q_2}\nonumber \\
    &+  \left[ \frac{10 ( t_\mathrm{TQG} + 2t_\mathrm{SQG})}{ T_{1,0}} \right] \left(  P_{|1\rangle}^{q_1} + P_{|1\rangle}^{q_2} \right). 
\end{align}
The first term in Eq.~\ref{eq:total_infidelity} and first line in Eq.~\ref{eq:total_infidelity_full} correspond to the errors of the two-qubit gate, the second term and line to single-qubit gate errors and the last to the state preparation error. The performance metric for the advanced noise model can be derived from Table~\ref{tab:noise_scaling}.

A similar fidelity approximation was defined in Refs. \cite{arute_2019_supremacy,google_calibration_2024}. We have introduced additional weights in the sum, in order to account for the relative number of single and two-qubit gates, and also to correctly take into account the fact that the error in state preparation occurs only once, while the gate error is significantly amplified after a number of applications of the operation. As the determination of the position of maximal $\mathcal{P}$ within the $(E_C,E_J)$ landscape, using which we find the optimal range for parameters, depends exclusively on the relative values of the weights $w_{i,j}$, we proceed by assuming that all error terms associated with the two-qubit gate are assigned weights of $w_\mathrm{TQG} = 1$. As pictured in Fig.~\ref{fig4:opt_EjEc}c, we analyze a circuit where we perform four single-qubit gates per each two-qubit gate, with half those single-qubit gates being $\pi$ rotations which are more susceptible to leakage. Note that this ratio of single to two-qubit gates arises naturally with the introduction of randomized compiling into the algorithm\cite{hashim_2021_rc}. The weights $w_{\mathrm{SQG},T_{1,\phi}}$ for the decoherence induced infidelity during a single-qubit gate therefore have a value of $2$ (since a $\pi$ and $\pi/2$ rotation are applied) per qubit, while the leakage error $w_{SQG,\mathrm{leak.}}  = 1$ per each qubit. Lastly, the thermal state preparation error is weighted so that $w_\mathrm{SP} = 10 ( t_\mathrm{TQG} + 2t_\mathrm{SQG}) / T_{1,0}$, since this quantity is the inverse of the approximate number of single and two-qubit gates that we can perform within a specified coherence time $T_{1,0}$ at the reference point. Therefore, the performance metric $\mathcal{P}$ approximates the fidelity of the simple circuit displayed in Fig.~\ref{fig4:opt_EjEc}c.

\begin{figure}[t]
\centering
\includegraphics[width=.8\textwidth]{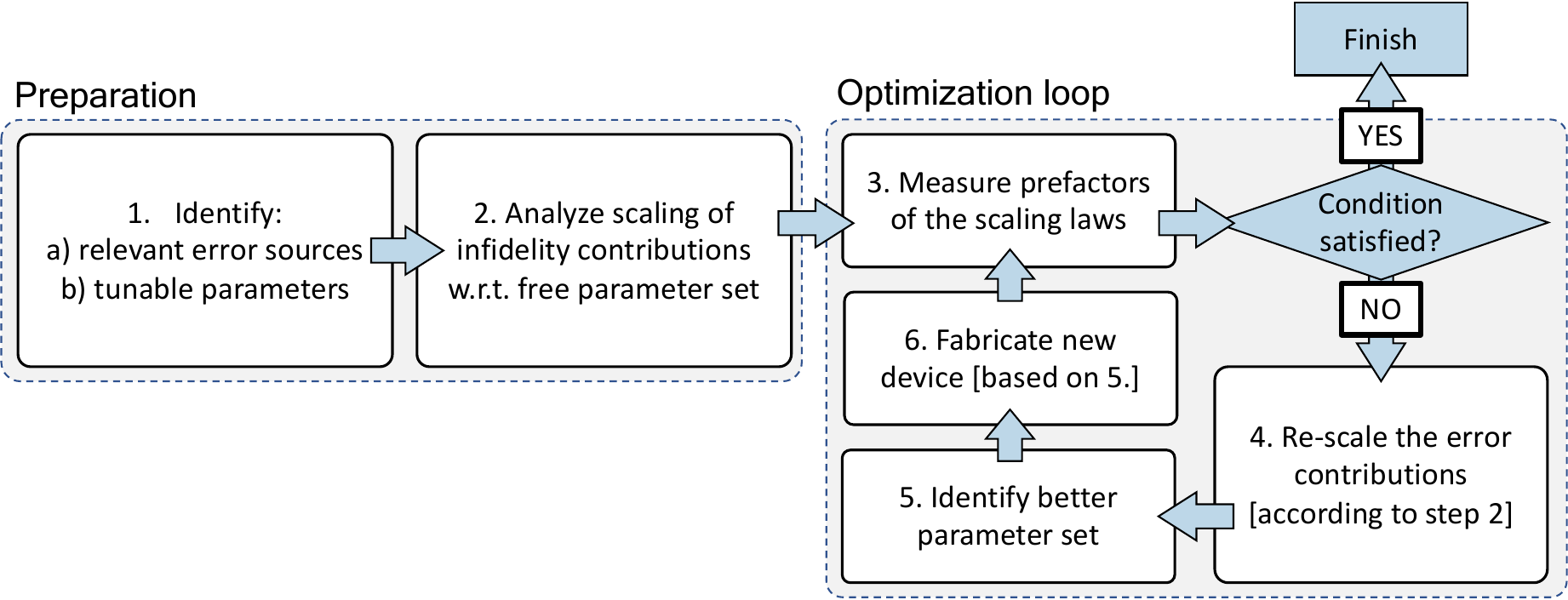}
\caption{\label{fig6:general_optimization} General quantum processor parameter optimization procedure. The procedure is composed of two parts: the preparatory phase and the actual optimization. The optimization loop is composed of first obtaining the necessary data to evaluate the performance metric from Eq.~\ref{eq:total_infidelity} (step 3), after which we can determine whether the value of $\mathcal{P}$ is good enough or whether the optimization should continue. In the latter case, the performance metric is extrapolated to different values of the quantum processor parameters (from step 1.b) we wish to tune as plotted in Fig.~\ref{fig4:opt_EjEc}d and a better parameter set is identified in step 5, based on the results from step 4. Afterwards, a new device is fabricated with the improved parameter set and the cycle repeats.  }
\end{figure} 

The values of $1 - \mathcal{P}$ from Eq.~\ref{eq:total_infidelity_full} for certain parameters are plotted in Fig.~\ref{fig4:opt_EjEc}d. Due to the fact that the maximum of $\mathcal{P}$ is an optimal solution only for the circuit pictured in Fig.~\ref{fig4:opt_EjEc}c, we have plotted an optimal region of parameters defined by the 10th percentile of the points with the lowest value of $1-\mathcal{P}$. This region is defined so that the parameters of the system can be further fine-tuned within the optimal area. This optimal region is also contrasted with the optimal region obtained with a density matrix simulation of the same circuit, and the good agreement between the curves shows that $\mathcal{P}$ is a valid performance metric. We also observe good alignment between the optimal regions of the two distinct considered noise models, with the second model anticipated to offer greater accuracy. The discrepancy is larger at lower coherence times, since both models differ mainly in the decoherence modeling of the diabatic CZ gate. Further numerical tests have shown that the more accurate infidelity formulas are the main reason for the difference\cite{abad2024}, rather than the increased flux-sensitivity of Qubit 1 during two-qubit gates, or the differing transmon idling frequency configurations.

The color scale corresponding to the $1 - \mathcal{P}$ has four distinct less-favorable regions, corresponding to the error sources in Table \ref{tab:noise_scaling}. For small $E_C$ and, consequently, low anharmonicity $\alpha_{q_2}$, the overall fidelities are low due to relatively large errors caused by leakage during single-qubit gates. For small $E_J$ and large $E_C$, $E_J/E_C$ has a low value and therefore the system experiences relatively large errors arising from parity switching events. If both $E_J$ and $E_C$ are large, the coherence times are short and, therefore, the dominating source of infidelity. Moving perpendicularly to the constant frequency contours towards lower values of $E_C$ and $E_J$, results in lower frequencies and, thus increased errors caused by thermal excitations.

On the other hand, when considering near-future transmons with coherence times close to 0.5 ms\cite{Wang_2022,place_2021} in Fig.~\ref{fig4:opt_EjEc}d, we observe that the optimal region is shifted towards larger values of $E_C$, and is even limited by the parity-switching effects in the bottom right corner, thus demonstrating the importance of this effect in future QPU design, when two-qubit gate infidelities surpass the limit of $10^{-3}$ infidelity.

By comparing the data in Fig.~\ref{fig4:opt_EjEc}b and d, we observe that the implementation by Xu, H., \textit{et al.} (2021)\cite{Xu_2021} is close to the parity-switching induced error region. Since the pure dephasing times of the system were not reported we are not able to assess the relative contribution of the parity switching error to this gate implementation. However, using the results shown in Eqs.
~\ref{eq:simplified_pert_cphase} and \ref{eq:cphase_Pee_infidelity}, we observe that the parity-switching-error induced infidelity for the parameters is estimated to be $1 - \mathcal{F} \approx 2\cdot10^{-3}$. As a comparison, the reported $\Gamma_1$ decay rates (measured in the idling configuration) in the reference were $1/\Gamma_1^{q_1} = 20.8\,\mu$s and $1/\Gamma_1^{q_2} = 28.8\,\mu$s\cite{Xu_2021}. Together with the reported gate time, these decay times correspond to an infidelity contribution of approximately $6\cdot10^{-4}$, as given by the expression in Table \ref{tab:noise_scaling}. These results show that for this specific implementation of the diabatic CZ gate, the described effects of the parity switches are comparable, and possibly even greater than the $T_1$ decay induced infidelity.

The shifts of the optimal region with increasing coherence times seen in Fig.~\ref{fig4:opt_EjEc}d mean that the qubit parameters must be adapted to the currently achievable coherence times. Fig.~\ref{fig5:example_opt_EjEc}a shows how the performance metric $\mathcal{P}$ increases with the achievable coherence properties of the system, provided that the parameters $E_J$ and $E_C$ are adapted to the coherence times. On the other hand, Fig.~\ref{fig5:example_opt_EjEc}b displays the value of the function $\mathcal{P}$ without changing $E_J$ and $E_C$. Although the performance metric $\mathcal{P}$ on panels a and b is initially largely coherence limited, the assumption of fixed parameters $E_J$ and $E_C$ in the simulations of Fig.~\ref{fig5:example_opt_EjEc}b shows that simply increasing the coherence times does not necessarily give better fidelities if the effects of other sources of error are not taken into account. In this case, the parity switching error becomes the dominant error source, as seen from the upper panel in Fig.~\ref{fig5:example_opt_EjEc}b.

It is also important to mention what other aspects of successfully operating a transmon tunable coupler based quantum computer were not included in the presented analysis.
\begin{itemize}
    \item We are neglecting any possible cross-talk effects between next-nearest neighbors.
    \item Idling errors were discussed in this work, but not included in this analysis, as their contribution depends significantly on the specific algorithm being implemented. Taking also this effect into account, the area with low $E_J/E_C$ is even less favorable.  
    \item Other sources of decoherence are expected to have smaller contributions and therefore do not significantly affect the findings presented here.
    \item The TLS environment is random, meaning that the scaling shown in Table \ref{tab:noise_scaling} only holds for the average of a large number of qubits. Additionally, the $T_1$ decay rate is also heavily influenced by the design, i.e., the exact geometry of the capacitor pads of the transmon, meaning that the presented results are only valid for a comparison of qubits with similar designs. 
    \item Recently, two-qubit gates with tunable couplers idled below the qubit frequencies have been demonstrated\cite{Marxer_2022,Sete_2021}. In such implementations, the presence of thermal coupler excitations limit the gate performance at lower qubit frequencies in a more complex manner. Our analysis in the low frequency regime therefore only holds for the original tunable coupler proposals, where the coupler is idled above the computational transmons.
\end{itemize}

While the results in Figs.~\ref{fig4:opt_EjEc}d and \ref{fig5:example_opt_EjEc}a,b demonstrate how the knowledge about the error contributions can be used to guide the design of superconducting circuits, the scheme can easily be generalized to any type of quantum processor and any set of error sources. The only requirement is that the infidelity contribution of each error source depending on the tunable parameters of the processor can be evaluated either analytically or numerically. Especially the latter is typically feasible for errors which are local, i.e. do not affect a large number of components.

The schematic in Fig.~\ref{fig6:general_optimization} shows how the optimization procedure can be generalized to different quantum processor parameters and errors. The preparatory phase includes first identifying the parameters of the quantum processor we wish to optimize as well as the error sources whose contributions depend on these parameters (step 1). In the transmon example this corresponds to $E_J$ and $E_C$ of the high-frequency qubit and the list of error sources in Table~\ref{tab:noise_scaling}. In the second step an analysis of the infidelity contributions of the selected error sources with respect to the tunable parameters is performed, as described in Table~\ref{tab:noise_scaling}. At this point, analytical formulas, such as the one in Eq.~\ref{eq:simplified_infidelity}, are invaluable. However, in certain cases numerical simulations are needed to obtain these relations. Generally speaking, numerical simulations are feasible in the case when the error source is local to a small number of qubits (i.e. the size of the Hilbert system for the simulation is not too large).

Once the preparatory phase is complete, the optimization step of the processor can commence by first obtaining the scaling coefficients needed to evaluate the infidelity contributions of each error (step 3 in Fig.~\ref{fig6:general_optimization}), and hence evaluate the performance metric $\mathcal{P}$. This step can include measurements of the average coherence times of the qubits, their effective temperatures, etc. In some cases, such as for the parity switching error described in this manuscript, no additional measurements are required. Once the performance metric is evaluated one can decide whether to proceed with the optimization based on a predefined condition which can be chosen arbitrarily. As an example, in Fig.~\ref{fig4:opt_EjEc}d, the convergence condition is satisfied if the current QPU parameters are within the optimal region defined by the top 10th percentile. If the value of the performance metric was found to be insufficient, a new set of parameters is chosen based on the the landscape of the performance metric (step 5), as shown in Fig.~\ref{fig4:opt_EjEc}d.

\section*{Discussion}

We have provided a novel framework for the optimization of circuit parameters that can be used to guide the future design of transmon-based quantum computers. Our findings reveal the presence of a distinct global performance peak within the $E_J$ and $E_C$ parameter space, which has not been identified before. Moreover, our optimization procedure can be straightforwardly extended to more error sources, provided that the scaling of the error as a function of the system parameters is known analytically, or the infidelity contribution can be evaluated numerically. The latter is typically realistic as long as the error is sufficiently local, i.e., it depends only on the parameters of a handful of transmons at most. While we have based our analysis on transmon qubits connected via tunable couplers, the same principles, albeit with different error sources, can be applied to different types of qubits\cite{Kjaergaard2020_review} or co-design chips\cite{algaba_2022}. Additionally, more parameters than just $E_J$ and $E_C$ can be optimized, e.g., also the gate durations can be considered as free parameters since they are realistically easy to adjust in experiments. The limiting factor here are the errors for which the analytical behavior is unknown and numerical interpolation in a large parameter space is too demanding.

We have chosen to present a range of optimal parameters, rather than prescribing a single optimal value for $E_J$ and $E_C$ since additional improvements of the noise models are expected to improve the accuracy and reliability of the procedure. An extension of the framework in the tunable coupler example would include also optimizing the $E_J$ and $E_C$ of the coupler transmon, as well as the couplings between the three transmons of the system. We are not aware of any analytical results linking the gate infidelity with these parameters, however it is known that the coupler coherence affects the effective qubit decoherence due to the hybridization of the states\cite{chu_2021,papic2023error}. Correctly taking into account the effects of different coupling strengths necessitates a good understanding of the effect of the state hybridization on single-qubit rotations\cite{heunisch2023tunable,marxer_aps_2024} and the two-qubit gate duration (Eq.~\ref{eq:simplified_pert_cphase}). We believe that the inclusion of pulse-level numerical modeling, as presented in Ref. \cite{papic2023error}, can also take into account such effects. Another straightforward improvement of our framework would be to consider a set of (possibly random) parameters for $E_J$ and $E_C$ in the evaluation of the performance metric in order to better mimic the spread in transmon frequencies required to avoid unwanted resonances in a more realistic lattice\cite{berke_2022}. This would also allow us to better estimate the effects of flux noise.

It is important to acknowledge the potential influence of other noise mechanisms, not explicitly addressed in this study, on the optimal design parameters. Specifically for superconducting circuits, the list of error sources in Table~\ref{tab:noise_scaling} can be expanded to include more contributions to the circuit coherence times, such as losses due to charge-coupling to an impedance, current fluctuations in the flux-bias lines, etc. All of these contributions to the coherence times can be evaluated using open-source software such as \verb|scqubits|\cite{Groszkowski_2021_scqubits}. We believe that even taking into account cross-talk effects is possible, using similar approaches as in Ref.~\cite{google_calibration_2024}. Nonetheless, the results in Fig.~\ref{fig4:opt_EjEc}d show that the optimal parameter regimes presented are robust to small perturbations in the idling configuration, and higher-order corrections to the noise models. Our findings, as depicted in Fig.~\ref{fig4:opt_EjEc}d, reveal that a two-fold variation in the reference coherence time only marginally adjusts the optimal parameter domain. This observation underscores the robustness of our results, suggesting that the presence of additional, potentially sub-leading noise mechanisms omitted from our simulations are unlikely to precipitate a drastic alteration in the presented outcomes.

Additionally, we have established how parity switching affects the commonly implemented tunable-coupler mediated diabatic and adiabatic CZ gate in a transmon based quantum computer, both analytically and numerically. We have shown that the parity switching error can be the main quasiparticle-related error source of the two-qubit gate. Moreover, we have demonstrated that the experimental implementation of the gate presented in Ref.\cite{Xu_2021} may have a comparable, if not larger, contribution of parity-switching errors compared to all $T_1$ decay mechanisms.

While the tunable-coupler-based diabatic and adiabatic CPHASE gates are more relevant due to its implementation in leading large-scale experiments\cite{arute_2019_supremacy,google_qec_2023,cao_2023,wu_2021}, we believe that the effects described in this manuscript should be considered in any current or future transmon-based quantum gate which utilizes higher-excited states, such as the gate schemes proposed in Refs.~\cite{Caldwell_2018,Sete_2021,Ghosh_2013}. Accordingly, we do not believe that the parity-switching effects play a major role in implementing iSWAP-like interactions in the tunable coupler architecture, since the states involved in the interaction are not significantly hybridized with the second excited state of any transmon. Furthermore, the effects on parity-switches on more general fSIM gates\cite{Foxen_2020} are expected to be of a similar magnitude as in the adiabatic CZ case presented in Fig.~\ref{fig3:idling}c.

One of the primary anticipated advantages of incorporating tunable couplers into the system is the potential for on-demand complete suppression of $ZZ$-type interactions among the qubits. However, our research in this context has revealed that the stochastic nature of parity switches imposes constraints on this proposition, practically establishing a lower bound on the achievable minimum $ZZ$ coupling strength. The magnitude of this "always on" interaction should be an important consideration when running longer algorithms. More specifically, this effect becomes relevant if the algorithm is long enough to accumulate a considerable conditional phase due to the unwanted coupling strengths shown in Fig.~\ref{fig3:idling}b. Since current coherence times are approaching the 1 ms limit\cite{Wang_2022,place_2021}, the residual idling strength can become relevant if the described effects are not taken into account in the design of the transmon parameters. 

One way of mitigating the parity-switching effects would be to attempt to tune the offset charge $n_g$ to the point where both parity manifolds are degenerate. However, such a solution is not practical, since the environmental charge noise would result in a drift of the offset charge $n_g$ as was demonstrated in Refs. \cite{christensen_2019,Wilen_2021}. We further note that $\cos(2\pi n_g) = 0$ is the low-frequency charge noise hotspot \cite{Koch_2007} in which the qubit frequency is maximally sensitive to the fluctuations of the offset charge. Therefore, the qubit is expected to have lower coherence times at this particular value for $n_g$.

\section*{Methods}

\subsection*{Modelling the Tunable Coupler Circuit and Diabatic $\mathrm{CPHASE}$ Gate}

The dependence of the transmon parameters on the parity and offset charge is not explicitly shown in Eq.~\ref{eq:tqg_ham} in order to simplify the notation. However, we note that the whole system has $2^3 = 8$ distinct parity states.
We emphasize also that the couplings between the transmons in Eq.~\ref{eq:tqg_ham}, $g_{ij}  = \beta_{ij} \sqrt{\omega_i \omega_j}$ also depend on the frequencies, meaning that while $g_{q_1 q_2}$ is constant, $g_{q_{1,2} c}$ is time dependent. The dimensionless prefactors $\beta_{ij}$ depend on the coupling capacitances, as well as self-capacitances of the transmons in the lumped-element circuit model\cite{Yan_2018}.

The computational basis in this scheme is formed by the \textit{eigenstates} of the Hamiltonian from Eq.~\ref{eq:tqg_ham} in the idling configuration (defined below), rather than the local (uncoupled) transmon states \cite{chu_2021}. Since the couplings act only as a perturbation to the uncoupled states, we identify the full Hamiltonian eigenstates corresponding to the uncoupled states. More specifically, the computational state $|ij\rangle$, $i,j \in \{0,1\}$ is the eigenstate $|\psi\rangle$ of the Hamiltonian in Eq.~\ref{eq:tqg_ham} with the maximal overlap $|\langle \psi |i_{q_1} 0_c j_{q_2} \rangle|$. 
This notation is employed throughout this manuscript and we typically omit the subscript indices $q_{1,2}$ and $c$. The kets with three indices (e.g. $|ijk\rangle$) always denote the local (uncoupled) Fock states of the three-transmon system. The kets with only two indices (e.g. $|ij\rangle$) are used to denote the \textit{eigenstates} of the whole system that are closest to the local (uncoupled) state $|i0j\rangle$. We denote with $\omega_{ij}$ the angular frequency of the computational state $|ij\rangle$. These eigenenergies of these states are also used to computed the ZZ interaction strength $\zeta_\mathrm{ZZ}$, as defined in Eq.~\ref{eq:zz_interaction}.

However, since the coupler frequency is tunable, it is possible to find one or two frequencies $\omega_c$ for which $\zeta_{\rm ZZ}=0$\cite{Yan_2018,Sung_2021,Marxer_2022}, provided that the qubit-qubit detuning $\omega_{q_1} - \omega_{q_2} \in [\alpha_{q_2},-\alpha_{q_1}]$. These special frequencies are referred to as the coupler idling frequencies and denoted with $\omega_c^\mathrm{idle}$.

A variety of pulse shapes can be used to implement the gate effectively and without inducing too many unwanted transitions\cite{chu_2021,Sung_2021,Rol_2019}. Even though our analytical analysis makes minimal assumptions about the pulse shape, we need to choose a specific form for the numerical simulations. In our simulations we use the flattop Gaussian pulse described by the formula\cite{Sung_2021,collodo_2020} 
\begin{equation}\label{eq:flattop_gaussian}
    f(t) = \frac{1}{2}\left[\mathrm{erf}\left(\frac{t - \tau_b}{\sqrt{2}\sigma}\right) - \mathrm{erf}\left(\frac{t - \tau_b  - \tau_c}{\sqrt{2}\sigma}\right) \right] - C.
\end{equation}
The flattop Gaussian is obtained by a convolution of a step function with duration $\tau_c$ and a Gaussian with parameter $\sigma$. The reasoning behind this choice is that the convolution of the flattop pulse with a Gaussian strongly suppresses the spectral component of the flattop pulse at higher frequencies, thus reducing the probability for unwanted transitions. An additional rise time of $\tau_b$ is also introduced, which we fix to $\tau_b = 2\sqrt{2}\sigma$. Since any gate must have a finite duration, we introduce a cut-off at time $T = 2\tau_b + \tau_c$. The constant $C$ is then introduced to correct for the discontinuity at the beginning and the end of a pulse with finite duration. The coupler frequency is varied accordingly $\omega_c(t) = \omega_c^\mathrm{idle} - Af(t)$. More details about the numerical simulations are given in Ref. \cite{papic2023error}. The parity of the whole system is then switched from one state to the other according to the charge dispersion given by Eqs.~\ref{eq:parity_difference} and \ref{eq:asymptotic_charge_dipersion}.

Since in our analytical approach we approximate the realistic pulse shape shown in Eq.~\ref{eq:flattop_gaussian} with a square pulse (i.e., we neglect any dynamics during the ramping up and down of the flux pulse), we have to be careful when comparing the results to those obtained with the numerical data. Particularly, the pulse duration $T$ in the numerics is in general different than the gate duration $t_{g}$ we have defined in our analytical derivation, since the full duration $T$ also includes the time needed to ramp the pulse up and down. When we compare our analytical results with numerical simulations, we first extract the number $n$ of Rabi oscillations from the simulated data by monitoring the population of the $|002\rangle$ state during the gate operation. After $n$ is determined, the analytical gate time is adjusted such that $t_g = n\pi/\Omega$. The effective gate time obtained in this manner does not differ significantly from the duration of the flat part of the pulse, typically less than 5 ns.

\subsection*{The Schrieffer-Wolff Transformation}\label{app:SWT}
The Schrieffer-Wolff (SW) transformation used in this manuscript was first introduced in Ref. \cite{Bravyi_2011} and a similar transformation has been applied to the computational subspace of the two-qubit system in Refs.\cite{Yan_2018,heunisch2023tunable}. The aim of the SW transformation in our case is not to diagonalize the system, but rather to decouple the coupler states from the computational transmons, thus enabling us to only study the reduced system, i.e. we want to simplify the full Hamiltonian into a more tractable reduced model, containing only the relevant states (i.e. the states which have a significant population).

The reason for this is the fact that the computational basis of the system is defined by the eigenstates of the system which have the maximum overlap with states of the form $|i_{q_1} 0_c j_{q_2}\rangle$, i.e. with the coupler always in the ground state. Any excitations of the coupler therefore lead to errors, so in order to analyze the ideal gate dynamics, we constrain ourselves only to the energy levels of the computational subspace and the second excited state used in the Rabi oscillation during the CPHASE gate operation.

In general, a SW transformation is obtained by noting that any unitary operator can be written as $\hat{U} = e^{\hat{S}} = \mathds{1} + \hat{S} + \frac{1}{2}\hat{S}^2 + \dots$, where $\hat{S}$ is anti-hermitian, $\hat{S} = -\hat{S}^\dagger$. Consequently a unitary transformation of an arbitrary Hamiltonian $\hat{H}$ can be expanded in terms of $\hat{S}$ as
\begin{equation}\label{eq:basic_SW_transform}
    \hat{U}\hat{H}\hat{U}^\dagger = \hat{H} + \left[ \hat{S},\hat{H}\right] + \frac{1}{2}\left[\hat{S}, \left[\hat{S},\hat{H}\right]\right] + \mathcal{O}(\hat{S}^3).
\end{equation}
As is typical in perturbation theory, we introduce the index $\alpha$ for bookkeeping purposes, and split the full Hamiltonian into a diagonal part and two off-diagonal perturbations, so that $\hat{H} = \hat{H}_0 +  \alpha\hat{V}_1+  \alpha^2\hat{V}_2$. Additionally we rewrite the operator $\hat{S}$ as a first order operator $\hat{S} \rightarrow \alpha\hat{S}$, since if the perturbation is small, $\hat{U}$ should be close to identity. In our case, $\hat{H}$ is given in Eq.~\ref{eq:tqg_ham} and the first-order perturbation $\hat{V}_1 = -\sum_{i = q_1,q_2} \hbar g_{ic}(\hat{a}_i^\dagger - \hat{a}_i)(\hat{a}_c^\dagger - \hat{a}_c)$ corresponds to the capacitive couplings of the two transmons (Qubits 1 and 2, $q_{1,2}$) to the coupler ($c$), the direct coupling between the qubits $\hat{V}_2 = - \hbar g_{q_1 q_2}(\hat{a}_{q_1}^\dagger - \hat{a}_{q_1})(\hat{a}_{q_2}^\dagger - \hat{a}_{q_2})$ is a second order perturbation, while $\hat{H}_0$ is a sum of the three independent anharmonic oscillator Hamiltonians. This hierarchy is chosen due to the fact that in all practical scenarios $g_{q_1 q_2} \ll g_{q_1 c}, g_{q_2 c}$\cite{Yan_2018}.

By plugging the ansaetze into Eq.~\ref{eq:basic_SW_transform}, and grouping the terms with the same order of $\alpha$, we obtain
\begin{equation}
    \hat{U}\hat{H}\hat{U}^\dagger = \hat{H}_0 + \alpha \left( \hat{V}_1 + \left[ \hat{S},\hat{H}_0  \right]\right) + \alpha^2 \left(\hat{V}_2 + \left[ \hat{S},\hat{V}_1\right] +\frac{1}{2}\left[ \hat{S} ,\left[\hat{S}, \hat{H}_0 \right]\right] \right) + \mathcal{O}(\alpha^3).
\end{equation}
Looking at the first-order term, it is natural to choose $\hat{S}$ such that $\left[ \hat{S},\hat{H}_0\right] = -\hat{V}_1$, i.e. so that we cancel any couplings to the coupler states up to lowest order. However in order to do so and account for the couplings to the higher state correctly, we generalize the transformation from Refs. \cite{Yan_2018,heunisch2023tunable} 
\begin{align}
    \hat{S}_i &= \sum_{n_{q_i},n_c \in \{0,1\} } \sqrt{(n_{q_i} + 1)(n_c + 1)} \left[ \frac{g_{q_ ic}}{\Delta_{q_i c} + n_{q_i} \alpha_{q_i} - n_c \alpha_c }\left( \hat{\pi}_{q_i}^{n_{q_i}+1,n_{q_i}} \hat{\pi}_{c}^{n_{c},n_{c}+1} - \hat{\pi}_{q_i}^{n_{q_i},n_{q_i}+1} \hat{\pi}_{c}^{n_{c}+1,n_{c}}\right)  \right.  \nonumber \\
    &- \left. \frac{g_{q_i c}}{\Sigma_{q_i c} + n_{q_i} \alpha_{q_i}  + n_c \alpha_c}\left( \hat{\pi}_{q_i}^{n_{q_i}+1,n_{q_i}} \hat{\pi}_{c}^{n_{c}+1,n_{c}} - \hat{\pi}_{q_i}^{n_{q_i},n_{q_i}+1} \hat{\pi}_{c}^{n_{c},n_{c}+1}\right)\right] , \\
    \hat{S} &= \hat{S}_1 + \hat{S}_2.
\end{align}
We have additionally defined $\Delta_{q_ic} = \omega_{q_i} - \omega_c $, $\Sigma_{q_i c} = \omega_{q_i} + \omega_c $ and the operators $\hat{\pi}_{k}^{n,m} = |n\rangle \langle m |$, acting in the Hilbert space of $k\in \{ q_1,c,q_2\}$.

Since we have assumed the coupler remains in the ground state at all times, the effective Hamiltonian is defined on the set of states $\{|0_{q_1} 0_c 0_{q_2}\rangle,|0_{q_1} 0_c 1_{q_2}\rangle,|1_{q_1} 0_c 0_{q_2}\rangle,|1_{q_1} 0_c 1_{q_2}\rangle,|0_{q_1} 0_c 2_{q_2}\rangle \}$. Additionally, we neglect any couplings outside of the subspace of interest, however the resulting effective Hamiltonian still contains terms coupling the levels $|000\rangle \leftrightarrow |101\rangle$ and $|000\rangle \leftrightarrow |002\rangle$. These couplings are neglected in the rotating-wave approximation, as these transitions do not conserve the total occupation number.

By additionally setting the energy of the ground state to zero we arrive at the effective subspace Hamiltonian from Eq.~\ref{eq:effective_hamiltonian}. The perturbative parameter values are given by
\begin{align}
    \tilde{\omega}_{q_i} &= \omega_{q_i}  + \frac{g_{q_i c}^2}{\Delta_{q_i c}} + \frac{2g_{q_i c}^2}{\Sigma_{q_i c} + \alpha_{q_i} } + \frac{g_{q_i c}^2}{\Sigma_{q_i c}},\label{eq:perturbative_hamiltonian_parameters_omega}\\
    \tilde{\alpha}_{q_i} &= \alpha_{q_i} -\frac{2g_{q_i c}^2}{\Delta_{q_i c} } + \frac{2g_{q_i c}^2}{\Delta_{q_i c} + \alpha_{q_i} } +\frac{4g_{q_i c}^2}{\Sigma_{q_i c} + \alpha_{q_i}} + \frac{g_{q_i c}^2}{\Sigma_{q_i c}} - \frac{3g_{q_i c}^2}{\Sigma_{q_i c} + 2\alpha_{q_i}}, \\
    \tilde{g}_{01,10} &= g_{q_1 q_2} + \frac{ g_{q_1 c}  g_{q_2 c} }{2} \left(\frac{1}{\Delta_{q_1 c}} + \frac{1}{\Delta_{q_2 c}} 
    -  \frac{1}{\Sigma_{q_1 c}} - \frac{1}{\Sigma_{q_2 c}} \right), \\
    \tilde{g}_{11,02} &= \sqrt{2}\left[ g_{q_1 q_2} + \frac{ g_{q_1 c}  g_{q_2 c} }{2} \left(\frac{1}{\Delta_{q_1 c}} + \frac{1}{\Delta_{q_2 c} + \alpha_{q_2}} 
    -  \frac{1}{\Sigma_{q_1 c}} - \frac{1}{\Sigma_{q_2 c} + \alpha_{q_2}} \right) \right]. \label{eq:perturbative_hamiltonian_parameters_g1102}
\end{align}

\subsection*{Deriving the Effective Diabatic Gate Unitary}

We consider here the truncated Hilbert space spanned by the states $\{|00\rangle, |01\rangle,|10\rangle,|11\rangle,|02\rangle\}$. In our analytical  considerations, we thus neglect the possibility of leakage, which is a good assumption for high-fidelity gates. We emphasize that the leakage effects are included in our numerical simulations which have been made with the full Hamiltonian defined in Eq.~\ref{eq:tqg_ham}. 

We additionally neglect the counter-rotating terms within the truncated Hilbert space and, after applying the Schrieffer-Wolff trnasofrmation to the Hamiltonian in Eq.~\ref{eq:tqg_ham}, we obtain
\begin{equation}\label{eq:effective_hamiltonian}
\hat{H}_\mathrm{eff}/\hbar \hat{=} 
    \begin{blockarray}{cccccc}
    |00\rangle & |01\rangle & |10\rangle & |11\rangle & |02\rangle \\
    \begin{block}{(ccc|cc)c}
      0 & 0 & 0 & 0 & 0 &\, |00\rangle \\
      0 & \tilde{\omega}_{q_2} & \tilde{g}_{01,10} & 0 & 0 &\, |01\rangle \\
      0 &  \tilde{g}_{01,10} & \tilde{\omega}_{q_1} & 0 & 0 &\, |10\rangle \\
    \cline{1-5}
      0 & 0 & 0 & \tilde{\omega}_{q_1} + \tilde{\omega}_{q_2} & \tilde{g}_{11,02} &\, |11\rangle \\
      0 & 0 & 0 & \tilde{g}_{11,02} &  2\tilde{\omega}_{q_2} + \tilde{\alpha}_{q_2} &\, |02\rangle \\
    \end{block}
    \end{blockarray},
\end{equation}
where the parameters with the tilde denote the perturbed parameters of the original full Hamiltonian from Eq.~\ref{eq:tqg_ham} (see Eqs.~\ref{eq:perturbative_hamiltonian_parameters_omega}-\ref{eq:perturbative_hamiltonian_parameters_g1102}) up to second order in $g_{q_i c}/(\omega_{q_i} - \omega_{c})$. Without any loss of generality, we have also assumed that the second excited state of the \textit{second} qubit ($q_2$) is used to perform the Rabi oscillation.

The pulse shape (see Eq.~\ref{eq:flattop_gaussian}) used to implement a diabatic CPHASE gate can typically be divided into three distinct stages: (i) fast sweep of the coupler frequency to an operation value close to the resonance with the qubits; (ii) long constant-frequency plateau at the operation frequency; (iii) fast sweep back to idling frequency of the coupler. 
Since the plateau is typically longer compared to the fast sweeps \cite{Sung_2021}, and the effective coupling strengths $\tilde{g}_{01,10}$ and $\tilde{g}_{11,02}$ are significantly smaller if the coupler is further detuned, we assume that the plateau stage is the only contributor to the dynamics and the effective Hamiltonian from Eq.~\ref{eq:effective_hamiltonian} is constant in time. We show later that this approximation is valid for realistic gate durations close to 40 ns and longer. If this approximation cannot be made, the form of the effective Hamiltonian remains unchanged. However, the analytical results derived in the remainder of this section become more complex due to the time dependence of the effective Hamiltonian.

The effective Hamiltonian in Eq.~\ref{eq:effective_hamiltonian} is block-diagonal. Using matrix exponentiation, one can readily obtain an effective unitary time-evolution operator. Assuming that the computational transmons are detuned by approximately one anharmonicity, i.e. $\tilde{\omega}_{q_1} -\tilde{\omega}_{q_2} \approx \tilde{\alpha}_{q_2}$, the amplitude of the Rabi oscillation between the single-excitation states $|01\rangle$ and $|10\rangle$ is lower than that of the transition between $|11\rangle$ and $|02\rangle$ because $|\tilde{\omega}_{q_1} -\tilde{\omega}_{q_2}|\gg \tilde{g}_{01,10}$. Thus, we can neglect the interaction between the single-excitation states.

Since we are interested in the operator acting on the computational subspace of the system, we further truncate the subspace by excluding the non-computational $|02\rangle$ state. After also accounting for the single-qubit phases, which is typically done via virtual $Z$-rotations \cite{mckay_2017}, we obtain the effective time-evolution operator $\hat{U}(t)$ from Eq.~\ref{eq:effective_unitary}. Eq.~\ref{eq:perturbative_P11} for $P_{11}$ is derived by recognizing the block-diagonal structure of the effective Hamiltonian in Eq. \ref{eq:effective_hamiltonian}. This simplifies the analysis, reducing it to a standard two-level Rabi oscillation\cite{nielsen00}. Eq.~\ref{eq:perturbative_cphase} is obtained by subtracting the phases of the single-excitation states from the phase of the $|11\rangle$ state.

\subsection*{List of Assumptions}

Here we summarize all of the approximations involved in deriving Eqs.~\ref{eq:simplified_pert_cphase} and \ref{eq:simplified_pert_P11} in the main text. These expressions are valid under the following assumptions:
\begin{enumerate}
    \item The initial assumptions used to derive the effective Hamiltonian in Eq.~\ref{eq:effective_hamiltonian} are valid, meaning that:
    \begin{enumerate}
        \item The second order perturbation theory used for the Schrieffer-Wolff transformation is valid, i.e. $g_{q_2c}^2/(\omega_{q_2} - \omega_c)^2 \ll 1$.
        \item The gate has low leakage outside of the considered subspace, spanned by the computational states and second excited state of the relevant qubit.
        \item The rotating wave approximation for the counter-rotating coupling terms is justified.
    \end{enumerate}
    \item The Rabi oscillation between the states $|01\rangle \leftrightarrow |10\rangle$ is negligible compared to the Rabi oscillation between $|11\rangle \leftrightarrow |02\rangle$. In the perturbative regime this is fulfilled if $\sqrt{\tilde{\Delta}^2 + 4\tilde{g}_{01,10}^2} \ll \Omega$ and $\tilde{g}_{01,10}^2/(\tilde{\Delta}^2 + 4\tilde{g}_{01,10}^2) \ll \tilde{g}_{11,02}^2/\Omega^2 $.
    \item The coupler frequency in the interaction regime is relatively constant. If this is not the case, the time dependency of the perturbative parameters must be taken into account.
    \item The gate has low leakage outside of the computational subspace.
    \item The simplified formula in Eq.~\ref{eq:simplified_pert_cphase} additionally neglects the terms proportional to  $g_{q_1 c}g_{q_2 c}/(\Delta_{q_2 c} + \alpha_{q_2})^3$ and $g^2_{q_2 c}/(\Delta_{q_2 c} + \alpha_{q_2})^3$, and smaller.
    \item We have assumed that the main contributor to the perturbation is the Qubit 2 ($q_2$) whose second excited state is populated during the gate operation. However, if the system is designed in such a way that the charge dispersion of any of the other two transmons is significantly larger, their effects might not be negligible anymore. 
\end{enumerate}
Note that both assumptions about the leakage are automatically fulfilled if the gate has a high-fidelity.

\subsection*{Fidelity Scaling}\label{app:fidelity_scaling}
Scaling of the fidelity with the number of gates, i.e. computing the average gate fidelity of sequential application of $N$ gates, can be found by using Eq.~\ref{eq:kraus_operators} to first define the map corresponding to a series of gates
\begin{equation}\label{eq:N_gate_kraus_map}
    \left(\mathrm{CPHASE}_\mathrm{CPS}\right)^N[\hat{\rho}] = \sum_{k=0}^N {{N}\choose{k}} \hat{U}_-^k \hat{U}_+^{(N-k)}\hat{\rho} \, \hat{U}_-^{(N-k)} \hat{U}_+^k,  
\end{equation}
where we have used the fact that the operators in Eq.~\ref{eq:kraus_op_definition} are diagonal and therefore commute with each other and also $\hat{U}_+^\dagger = \hat{U}_-$. 
Using the effective Kraus operators from Eq.~\ref{eq:N_gate_kraus_map}, and only considering the conditional phase error, we combine this with the fidelity definition from Eq.~\ref{eq:fidelity_definition}, and obtain
\begin{align}\label{eq:infidelity_scaling}
     \mathcal{F}[\left(\mathrm{CPHASE}_\mathrm{CPS}\right)^N] &=  \frac{d + \sum_{k = 0}^N {{N}\choose{k}}  \left|\mathrm{tr}\left\{\left(\hat{U}_\mathrm{CPHASE}^\dagger\right)^N\hat{U}_-^k \hat{U}_+^{(N-k)}\right\}\right|^2}{d^2 + d}\\
     &= \frac{4 + \frac{1}{2^N}\sum_{k = 0}^N {{N}\choose{k}} \left[10 + 6\cos\left( \frac{\delta\phi}{2}(N - 2k)\right)\right]}{20}\\
     &\approx 1 - \frac{3 }{80}(\delta\phi)^2 \frac{1}{2^N}\sum_{k=0}^N {{N}\choose{k}} (N - 2k)^2 \\
     &= 1 - \frac{3}{80} N (\delta\phi)^2.
\end{align}
This result indicates that calibrating the gate such that the error is evenly split between the two parities not only increases the single-gate fidelity but also leads to a more generous scaling of the infidelity $\propto N$, compared to the purely coherent error case for which the error scales as $\propto N^2\,$\cite{hashim_2021_rc}.

\subsection*{Gate parameters for high-fidelity simulations}\label{app:gate_params}

Finding good gate parameters, both for the Hamiltonian as well as for the pulse for high-fidelity simulations, is not a trivial task. Here, we discuss how to find optimal Hamiltonian parameters at different qubit frequencies and anharamonicities. In general, the only prerequisite is that the qubits are detuned by approximately one anharmonicity, which can be seen from Eq.~\ref{eq:effective_hamiltonian}.

As seen from Table \ref{tab:tqg_pert_parameters}, we keep some of the parameters in the simulation fixed, while others depend on $\alpha_{q_2}$, which is varied. These parameters are based on the experimental values from Ref. \cite{Sung_2021} and the coupling coefficients from the Hamiltonian in Eq.~\ref{eq:tqg_ham} are related to $\beta_{ij}$ via the following relation $g_{ij} = \beta_{ij}\sqrt{\omega_i \omega_j}$, as described in the main text. The small perturbation to $\alpha_{q_1}$ is there in order to ensure that the energy levels are significantly non-degenerate for perturbation theory to apply. This is also completely realistic as a typical fabrication procedure results in seemingly random deviations from the designed values.
More specifically, the gates simulated in Fig.~\ref{fig2:gate_results}(a,b) are obtained by varying $\alpha_{q_2}$ with values $\alpha_{q_2} \in [195,230,250,270,300]\,h\,$MHz. Additionally, in order to certify that the analyzed effect is not limited to the choice of parameters presented above, the second data point of Fig.~\ref{fig2:gate_results}(a,b) at $E_J/E_C \approx 50$, was generated in the same way as in Table \ref{tab:tqg_pert_parameters}, but with the change to $\omega_{q_2} = 5.1\cdot 2\pi\mathrm{GHz}$. 
The coupler frequency in the idling configuration $\omega_c^\mathrm{idle}$, i.e. before and after performing a gate, is determined by diagonalizing the Hamiltonian to fulfill the condition $\zeta_\mathrm{ZZ} = 0$  from Eq.~\ref{eq:zz_interaction}.
The pulse parameters from Eq.~\ref{eq:flattop_gaussian} are obtained by numerically optimizing the fidelity of the gate, with fixed $\sigma = 5\,$ns and $\tau_b = 2\sqrt{2}\sigma$. Typical values of the amplitude are $A\sim 1 - 1.2 \cdot \,2\pi \,\mathrm{GHz}$ and $\tau_c \sim 60\,$ns. 

\subsection*{Scaling of Transmon Error Sources}\label{app:noise_scaling}
In this section we derive the scaling of the noise parameters shown in Table \ref{tab:noise_scaling} with the transmon energies $E_J$ and $E_C$. The noise is modeled by appending the appropriate noise channel after the gate unitary, and the calculated infidelity is thus independent of the unitary dynamics, as was shown in Ref. \cite{abad_2022}.

\subsubsection*{Charge noise $T_1$}
The $T_1$ decay time of transmon devices is believed to be currently limited by the presence of a discrete number of environmental two-level systems (TLSs) which couple to the qubit via their electrical dipole \cite{Muller_2019,siddiqi_2021_review,carroll_2022,Premkumar2021,cho2022,Lisenfeld_2019,abdurakhimov_2022}.
Therefore assuming that the charge noise in the system is weak enough, the interaction Hamiltonian between a transmon and an environmental TLS can be derived from Eq.~\ref{eq:full_transmon_hamiltonian}, by replacing $n_g \rightarrow n_g + \delta\hat{n}_g$, as in Refs. \cite{Muller_2019,abdurakhimov_2022}, so that
\begin{equation}
    \hat{H}_\mathrm{q-TLS} = 8 E_C \, \hat{n}\otimes \delta\hat{n}_g = -4\sqrt{2} E_C\left( \frac{E_J}{8E_C}\right)^{\frac{1}{4}} i (\hat{a} - \hat{a}^\dagger)\otimes \delta\hat{n}_g. 
\end{equation}
In the above equation, we have used the asymptotic expression for the number operator $\hat{n}$, derived already in Ref. \cite{Koch_2007}. The operator $\delta\hat{n}_g$ is defined in the Hilbert space of the TLS, and is related to the parameters of the TLS - more specifically, its electrical dipole \cite{cho2022}. More importantly, $\delta\hat{n}_g$ does not explicitly depend on $E_J$ and $E_C$. 

Since the majority of the $T_1$ experiments on transmons display exponential decays\cite{Burnett_2019}, the dynamics can be captured by the Lindblad equation. As a second order approximation is assumed in the derivation of the Lindblad equation, the resulting decay rates are found to be proportional to the square of the coupling coefficient of the transmon to the TLS environment \cite{breuer_petruccione_oqs}. In our case this translates to 
\begin{equation}
    \Gamma_1 \propto E_C^{3/2} E_J^{1/2}.
\end{equation}
Here we have omitted the noise spectrum of the environment, since current models assume a flat noise spectrum without any dependence on the qubit frequency\cite{cho2022}. 

\subsubsection*{Flux noise $T_\phi$}

While it is not strictly necessary for the computational transmons to be flux-tunable, it is often desired as flux-tunability additionally enables the implementation of an iSWAP gate with the same architecture \cite{Sung_2021,Yan_2018}. Having slow flux tunability is also desirable in order to avoid resonances with TLSs in the environment which can severely limit the $T_1$ decay time \cite{Lisenfeld_2019,abdurakhimov_2022}, as well as helping with the issue of frequency crowding \cite{berke_2022}.

However, unlike the random noise spectrum of a TLS environment producing $T_1$ dynamics, the noise spectrum of magnetic-flux noise is typically observed to have a $1/f^\alpha$ frequency dependence \cite{rower_2023,Braumuller_2020,Yan_2016,siddiqi_2021_review}, with $ \alpha \sim 1$. The large noise-spectrum amplitude at lower frequencies means that the long-time correlation results in non-Markovian dynamics \cite{Papic_2022}. Assuming that the noise is slow enough, in order for the adiabatic approximation to hold, the interaction Hamiltonian due to a slowly fluctuating magnetic environment $\delta\hat{\Phi}$ is given by \cite{Koch_2007}
\begin{equation}
    \hat{H}_\mathrm{q-flux} = \frac{\partial \hat{H}}{\partial \Phi} \otimes \delta\hat{\Phi} = \hbar \frac{\partial \omega}{\partial \Phi} \hat{a}^\dagger\hat{a} \otimes \delta\hat{\Phi}.
\end{equation}
Similarly to the $T_1$-decay scenario, we make the assumption that the environment operator $\delta\hat{\Phi}$ remains independent of the transmon parameters. In reality, the magnitude of this operator is contingent upon the inner product between the magnetic dipole operator of the spins and the surface vector of the SQUID loop.

The form of the flux dispersion $\partial \omega / \partial \Phi$ of a split-junction transmon, with Josephson energies $E_{J_1}$ and $E_{J_2}$, is determined by the relations~\cite{Koch_2007}
\begin{align}
    E_J(\Phi) &= E_{J\Sigma}\cos(\pi\Phi)\sqrt{1 + d^2\tan^2(\pi\Phi)},\\
    \hbar\omega(\Phi) &= \sqrt{8 E_C E_J(\Phi)} - E_C,
\end{align}
with $E_{J\Sigma} = E_{J_1} + E_{J_2}$ and $d = |E_{J_1} - E_{J_2}|/(E_{J_1} + E_{J_2})$. We assume that $d \lesssim 1$, since values of $d \approx 1$ result in overall longer coherence times, and the computational transmons need to be tuned over a smaller frequency range, typically on the order of $\sim 100\,$MHz\cite{Sung_2021}. In this regime, the flux dispersion can be approximated as 
\begin{equation}\label{eq:flux_dipsersion}
    \frac{\partial \omega}{\partial \Phi} = (d-1)\sqrt{8E_C E_{J\Sigma}}\frac{\pi\sin(2\pi\Phi)}{2}  + \mathcal{O}([d-1]^2),
\end{equation}
and the decay rate scales with the coupling coefficient \cite{Bergli_2009,paladino_2014},
\begin{equation}
    \Gamma_\phi(\Phi) \propto \left|\frac{\partial \omega}{\partial \Phi} \right|.
\end{equation}
Ideally, the system is designed so that both the idling frequency of the computational qubits (i.e. the frequency at which single-qubit rotations are performed), as well as the computational transmon frequency during the two-qubit gate operation are positioned at the first-order flux-insensitive sweet-spot. In that case, it can be seen from Eq.~\ref{eq:flux_dipsersion}, that the flux noise decay rate is identical in both the idling configuration as during the gate operation. This is true for both computational qubits.

While we have explicitly addressed only the first-order flux dispersion $\partial\omega/\partial\Phi$, it is evident that the aforementioned conclusion remains valid even if $\partial\omega/\partial\Phi = 0$. In such cases, we must consider the second order dispersion $\partial^2\omega/\partial\Phi^2$. A simplified scaling relationship of $\Gamma_\phi$ with $E_C$ and $E_J$ can be derived under the condition that the applied flux $\Phi$ is independent of $E_J$ and $E_C$, or equivalently, when we are interested in the flux-averaged decay rate $\bar{\Gamma}\phi = \int_0^1 \mathrm{d}\Phi,\Gamma\phi(\Phi)$. This yields the relation:
\begin{equation}\label{eq:flux_noise_simple_scaling}
    \Gamma_\phi \propto \sqrt{E_C E_{J\Sigma}},
\end{equation}
which holds for all values of $d$.

The averaging over the flux is relevant, in a more realistic scenario, where the frequencies of the system may be calibrated in order to avoid resonances with individual TLSs and other elements on the chip\cite{google_calibration_2024}. Due to fabrication inaccuracies\cite{hertzberg_2021} and the stochastic nature of the TLSs, it is impossible to deterministically predict the optimal frequencies in an array of transmons, which prompts us to consider the flux-averaged decay rate scaling in Eq.~\ref{eq:flux_noise_simple_scaling}. 

Realistic gate implementations frequently involve fast flux tuning of one of the computational transmons to enhance gate fidelity \cite{Sung_2021,Foxen_2020}. This flux tuning influences the flux dispersion of the qubit and consequently impacts the decay rate during the two-qubit gate implementation. While predicting the exact detunings of all qubits proves challenging, in a more realistic noise model, we incorporate the flux dependence of the dephasing rate from Eq.~\ref{eq:flux_noise_simple_scaling} to better model the dephasing during the operation of one of the transmons in the two-qubit gate. Typically, tuning the low-frequency transmon ($q_1$) proves more advantageous, as the second excited state of Qubit 2 is inherently more susceptible to decoherence.

In the advanced noise model, we therefore consider that the frequency of Qubit 1 satisfies $\omega_{q_1} - \omega_{q_2} = \alpha_{q_2}/2\approx\alpha_{q_1}/2$ for all values of $E_J$ and $E_C$ during single-qubit gate operation. This will reduce any potential leakage due to driving the $|10\rangle \leftrightarrow |02\rangle$ transition when instead of $|10\rangle \leftrightarrow |11\rangle$\cite{marxer_aps_2024}. More specifically we consider Qubit 1 to be weakly tunable ($d=0.9$) and we assume its frequency is shifted so that the two-qubit gate resonance condition is satisfied, i.e. $\omega_{q_1} - \omega_{q_2} = \alpha_{q_2}$ during the two-qubit gate operation. Otherwise the transmons are assumed to be idling 10 MHz away from the sweet-spot. This information is sufficient to compute the decay rate of Qubit 1 at a different flux bias, by using Eq.~\ref{eq:flux_dipsersion}. We believe that pinpointing more specific detuning values would require pulse-level simulation.

\subsubsection*{Single-Qubit Gate Leakage}
The low anharmonicity of the transmon is a limiting factor in the operation of single-qubit gates \cite{motzoi_2009,papic2023error} as fast operations drive a part of the population from the computational subspace into the second excited state. A straightforward and effective scheme for mitigating this effect while implementing single-qubit gates known as Derivative Removal by Adiabatic Gate (DRAG) was presented in Ref. \cite{motzoi_2009}.

Since typical pulse amplitudes used to perform single-qubit operations are typically much lower compared to the qubit frequency, and the drive is resonant with the qubit, we assume that the rotating-wave approximation is accurate. In the frame rotating with the qubit frequency, the effective Hamiltonian depends on the pulse parameters, anharmonicity, and the detuning between the drive and qubit frequencies. This indicates that the amount of leakage does not explicitly depend on the qubit frequency. 

Thus, the average gate fidelity, similar to the one defined in Eq.~\ref{eq:fidelity_definition}, but in this case for a single-qubit rotation around the $x$ or $y$-axes, depends only on the qubit anharmonicity (or charging energy), since $\alpha \simeq -E_C$ in the transmon limit. This means that even though analytical results are not available, the relationship can be determined numerically and interpolated. This relationship, albeit with different parameters, has already been plotted in Ref. \cite{motzoi_2009}, and generally follows a dependence of $P_\mathrm{leak.} \propto E_C^{-\gamma}$, $5 \lesssim \gamma \lesssim 6$, with higher exponents observed at lower $E_C$. The independence of the single-qubit gate infidelity of the qubit frequency (within the transmon regime) was also verified numerically.

The single-qubit gate parameters assumed in Fig.~\ref{fig4:opt_EjEc}d and Fig~\ref{fig5:example_opt_EjEc} were a DRAG Gaussian pulse, with a $\sigma = 4\,$ns and a total duration $t_\mathrm{SQG} = 4\sigma$. The amplitudes of both DRAG components are numerically optimized before interpolating the dependence of $P_\mathrm{leak.}$ on $\alpha$, which was used to generate Fig.~\ref{fig4:opt_EjEc}d. The pulse drive frequency is assumed to be resonant with the qubit. More details are available in Ref. \cite{papic2023error}. We note here, that most of the SQG infidelity is due to leakage, rather than phase errors. 

\subsubsection*{Thermal Excitation Error}
While other error sources affect the gate performance, the thermal-excitation error considered here only affects the state preparation. 
The average gate fidelity, which is defined as the fidelity averaged over Haar random distributed input states, is therefore not applicable, since this error only affects one input state. We therefore replace the average gate infidelity \cite{Emerson_2005} with a state infidelity. This can also be qualitatively thought of as replacing the Haar random distribution with a delta-like distribution with a peak at the $|0\rangle$ state. By modeling the thermal excitation as a bit-flip channel $\mathcal{E}[\hat{\rho}] = (1 - P_{|1\rangle}) \hat{\rho} + P_{|1\rangle} \hat{\sigma}_x\hat{\rho}\hat{\sigma}_x$ it is straightforward to see that the state infidelity\cite{nielsen00}
\begin{equation}
   \mathcal{F} = \langle 0 |\mathcal{E}[|0\rangle \langle 0 |]  | 0\rangle = 1 - P_{|1\rangle} =  \frac{1}{1 + e^{-\beta \omega}},
\end{equation}
where we have additionally assumed that the temperature is low enough such that the population of the higher-excited states is negligible. 

\section*{Data availability}
The datasets used and analyzed in this paper are available from the corresponding authors upon reasonable request.

\section*{Code availability}
The code used to generate the data for this paper is available from the corresponding authors upon reasonable request.

\bibliography{biblio}

\section*{Acknowledgments}

We would like to thank Gianluigi Catelani and all of the employees at IQM for fruitful discussions, especially Johannes Heinsoo, Attila Geresdi, Antti Veps\"al\"ainen, Alejandro G\'omez Frieiro, Alessandro Landra, Fabian Marxer, Frank Deppe, Vladimir Milchakov, Hao Hsu, Manish Thapa and Hsiang-Sheng Ku. We would additionally like to acknowledge the support from the German Federal Ministry of Education and Research (BMBF) under Q-Exa (grant No. 13N16062), QSolid (grant No. 13N16161) and MUNIQC-SC (grant No. 13N16188).

\section*{Author contributions}
M.P. performed the analysis of the parity-switch effects and optimal parameter regime and wrote the draft of the manuscript. J.T. performed the analytical analysis of the tunable coupler system. A.A. and I. de V. supervised the work. A.H. conceptualized and supervised the project. All authors read, revised and approved the final manuscript.

\section*{Competing interests}
The authors declare the following competing interests: The transmon parameter design optimization procedure presented in this manuscript is part of the patent application number: EP 23193779.8.

\section*{Tables}

\begin{table}[h]
\centering
\caption{\label{tab:tqg_pert_parameters} Table of Hamiltonian parameters (Eq.~\ref{eq:tqg_ham}) with which high-fidelity CZ gates are possible.}
\begin{tabular}{c|c}
$\alpha_{q_1}$ & $\alpha_{q_2} + 10\cdot h\,\mathrm{MHz}$\\
$\alpha_{c}$ & $-110\cdot h\,\mathrm{MHz}$\\
\hline
$\omega_{q_2}$ & $4.8\cdot h\,\mathrm{GHz}$\\
$\omega_{q_1}$ & $\omega_{q_2} + \alpha_{q_2} + 10\cdot h\,\mathrm{MHz}$\\
\hline
$\beta_{q_1 c}$ & 0.015 \\
$\beta_{q_2 c}$ & 0.015 \\
$\beta_{q_1 q_2}$ & 0.001 \\
\end{tabular}
\end{table}

\begin{table}[ht]
\centering
\caption{\label{tab:noise_scaling} Scaling analysis for the most relevant error sources in an architecture of flux-tunable transmons, with two noise models. }
\begin{tabular}{ccccc}
& Parameter scaling  & Single-qubit gate  & Two-qubit gate infidelity & Two-qubit gate infidelity \\ 
&   & infidelity & (basic noise model) &(advanced noise model) \\ 
\hline \hline
Charge noise ($T_1$)  & $\Gamma_1 = \frac{1}{T_1} \propto E_C^{3/2}E_J^{1/2}$ & $ \frac{1}{3}\Gamma_1 t_\mathrm{SQG} $&  $ \frac{2}{5}(\Gamma_1^{q_1} + \Gamma_1^{q_2} ) t_\mathrm{TQG} $ &  $ \left(\frac{3}{10}\Gamma_1^{q_1} + \frac{1}{2}\Gamma_1^{q_2} \right) t_\mathrm{TQG} $ \\
Flux noise ($T_\phi$) & $\Gamma_\phi = \frac{1}{T_\phi} \propto E_C^{1/2} E_J^{1/2}$ & $ \frac{1}{6}\Gamma_\phi t_\mathrm{SQG} $ &  $ \frac{1}{5}(\Gamma_\phi^{q_1} + \Gamma_\phi^{q_2} ) t_\mathrm{TQG} $ & $ \left(\frac{3}{8}\Gamma_\phi^{q_1}(\Phi_\delta) + \frac{31}{40}\Gamma_\phi^{q_2} \right) t_\mathrm{TQG} $\\
Leakage & $P_\mathrm{leak.} \propto E_C^{-\gamma}$, $5 \lesssim \gamma \lesssim 6$ &  $\frac{1}{3}P_\mathrm{leak.} $ & / & / \\
Parity Switch & $(\delta\phi)^2 \propto e^{-2\sqrt{8 E_J / E_C}} E_C^{1/2} E_J^{7/2} $ & / & $\frac{3}{80}\left(\frac{t_\mathrm{TQG}}{2\hbar}\epsilon^{q_2}_2 \right)^2$ & $\frac{3}{80}\left(\frac{t_\mathrm{TQG}}{2\hbar}\epsilon^{q_2}_2 \right)^2$ \\
\hline
Thermal excitation & $P_{|1\rangle}/P_{|0\rangle} = e^{-\beta (\sqrt{8E_J E_C} - E_C)}$ & $P_{|1\rangle}$ & / & /
\end{tabular}

\end{table}

\end{document}